\documentclass[acmsmall]{acmart} 
\AtBeginDocument{%
  }

\setcopyright{cc}
\setcctype{by}
\acmJournal{PACMHCI}
\acmYear{2026} \acmVolume{10} \acmNumber{7} \acmArticle{GAMES065}
\acmMonth{11} \acmDOI{10.1145/3831289}

\received{February 2026}
\received[revised]{June 2026}
\received[accepted]{July 2026}

\usepackage{graphicx}
\usepackage{multirow}
\usepackage{placeins}
\usepackage{flafter}
\usepackage[skip=2pt]{caption}
\usepackage{amsmath}    
\usepackage{amscd}
\usepackage{float}
\usepackage{adjustbox}
\usepackage{physics}
\usepackage{rotating}   
\usepackage[toc,page]{appendix}
\usepackage{textcomp}   
\usepackage{changepage} 
\usepackage{setspace}
\usepackage{makecell}

\usepackage[flushleft]{threeparttable}
\usepackage{threeparttablex}
\usepackage{bbding}
\usepackage{tabularx}
\usepackage{lscape}      
\usepackage{booktabs}    
\usepackage{array}
\usepackage{lipsum}      
\usepackage{enumitem}    
\usepackage{ragged2e}
\usepackage{placeins}
\usepackage{subcaption}
\usepackage{tikz}

\begin{document}

\title[Strategic Gaze]{Strategic Gaze: Attention Allocation and Transition Patterns Across Functional Areas of Interest by Gameplay Outcome}

\author{Yufei Cao}
\email{yufei.cao1@anu.edu.au}
\orcid{0000-0002-9616-9913}
\author{Penny Sweetser}
\email{penny.kyburz@anu.edu.au}
\orcid{0000-0002-6543-557X}
\affiliation{%
  \institution{The Australian National University}
  \city{Canberra}
  \country{Australia}
}

\renewcommand{\shortauthors}{Cao et al.}

\begin{abstract}
Video games present players with complex, spatially distributed information across interface elements, with attention shaped by visual features and task goals. Eye tracking provides a useful method for examining player attention through gaze behaviour during gameplay. Yet empirical game research has relied on accumulated fixation measures that capture where attention is directed and how long it is maintained within regions, leaving less known about how gaze moves between regions to coordinate distributed information. We address this gap by integrating distribution-, duration-, and transition-based gaze measures across functionally organised interface regions in relation to gameplay outcomes. We conducted a within-subject study with 32 participants using a deck-building game, defining six functional Areas of Interest (AOIs) within the turn-based combat interface, spanning enemy, player, action, and auxiliary elements. We computed AOI hit, dwell time, transition probabilities, and entropy to compare gaze behaviour across outcome groups. Players in the win group showed more selective attention allocation extending to peripheral auxiliary resources, with more frequent action-oriented gaze transitions, a wider range of AOI pairs, and a more even distribution across AOIs. Within a strategic game setting, our findings show that gameplay outcomes are reflected in how players distribute, sustain, and shift visual attention across functional interface elements, offering a more comprehensive account of attention organisation in gameplay. 
\end{abstract}


\maketitle

\section{Introduction}
Visual attention regulates how information is selected and prioritised under limited cognitive resources \citep{kahneman1973attention, wickens1995multiple}.
In video games, visual attention operates at the game interface between perception and action, where players continuously perceive, interpret, and act on visual input embedded in the game world \citep{jorgensen2013gameworld}.
Within this process, how attention is distributed across the game display is closely tied to information processing and task performance, supporting effective action, reasoning, and communication during gameplay \citep{llanos2011players, sundstedt2013visual, roda2011human, cheng2014relationship}.
However, regulating visual attention in complex game environments can be demanding. Gameplay often requires players to track multiple information streams concurrently, with increasing visual load intensifying demands on attentional resources under time and performance constraints \citep{yu2024effects, Chung2017TheIO}. Beyond information quantity, attentional demands are also shaped by the functional heterogeneity of on-screen content. Game interfaces present different types of task-relevant information (e.g., navigational cues, status information), requiring players to regulate attention across these functionally distinct regions during task execution to support ongoing task goals \citep{jiang2019applying, almeida2016video}.

Eye tracking has been used as an evaluation method for examining how players visually interact with game interfaces and explore game environments \citep{almeida2011eyes, sundstedt2012gazing, zammitto2014gaming, mat2011eye}. By continuously tracking gaze location over time, eye tracking provides gaze data for characterising the allocation and temporal dynamics of visual attention \citep{Rayner1998, holmqvist2011eye, duchowski2017eye}. In particular, analyses of gaze behaviour have helped game researchers better understand which elements of the display attract users’ visual attention and for how long, and which elements are overlooked \citep{zammitto2014gaming}. By analysing the spatial distribution of gaze across game displays, prior game studies have shown that players’ prioritisation of gameplay information is influenced by both interface characteristics, such as colour and motion, and task goals \citep{el2006visual, sundstedt2012gazing}. In addition, prior work has related gaze behaviour to player related differences, showing that visual attention patterns can vary with different levels of player experience and performance. A review of findings on eye movement characteristics among eSports players with different expertise levels reported that expert players exhibit more targeted visual strategies and devote greater visual attention to key elements of the game \cite{luo2025differences}. For example, high-skill StarCraft players showed higher fixation ratios in task relevant AOIs and broader visual information acquisition during gameplay \cite{jeong2022difference}.

Eye tracking in game research is often challenged by the highly interactive nature of gameplay and the visually cluttered information presented on game interfaces \citep{sundstedt2012gazing, zammitto2014gaming}. Treating the whole display as a single undifferentiated visual field makes gaze based evaluation difficult to interpret, especially when players’ actions change what appears on screen during ongoing tasks \citep{zammitto2014gaming}. Clear units of analysis are therefore needed to interpret gaze data in relation to specific game elements and their gameplay roles. Accordingly, Area of Interest (AOI) methods have been used in game studies to analyse gaze behaviour by mapping gaze data onto predefined interface regions. AOIs refer to regions of interest within a stimulus where researchers collect gaze data to determine whether participants looked at expected areas and characterise eye movements within those areas \citep{holmqvist2011eye}. In game research, AOI methods have been used to examine how gaze is distributed across different game elements and how it varies across players. For example, \citet{conati2013understanding} defined a single AOI over an adaptive hint message and measured fixation time to examine students’ attention to the hint in relation to their performance and attitudes towards help in an educational game. Similarly examining player-level differences, \citet{wang2024comparative} compared eSports players’ attention to static FIFA screenshots using fixation counts within multiple expert-defined AOIs covering different regions of the game interface. Foundational eye tracking work has established a structured methodological basis for analysing AOI based gaze behaviour through multiple event types, examining both how gaze is distributed within interface regions and how it transitions between them \citep{holmqvist2011eye}. However, empirical applications in game research have largely concentrated on gaze distribution within regions, typically operationalised through accumulated fixation measures such as fixation counts and fixation durations \citep{wang2024comparative, conati2013understanding, lan2026impact, jeong2022difference}. Transition based measures, which capture gaze movement between AOIs and provide valuable insight into how spatially distributed information sources are coordinated during ongoing decision making \citep{goldberg1999computer}, have received less attention in game research. This gap is particularly relevant for game environments that involve a series of decision processes, where interfaces present multiple information sources that differ in their interactivity and functional roles, requiring players to continuously shift and organise attention between these elements during decision making. In addition, although AOI definitions vary across game studies, the procedures used to define and validate AOIs are rarely described in detail \citep{wang2024comparative, jeong2022difference, gotardi2019combining, joyce2024less, luo2025differences}. A recent systematic review identifies this inconsistency as a barrier to comparing gaze-based findings across studies \citep{luo2025differences}. To address these gaps, we examine player visual attention through a systematic functional AOI analysis in relation to performance differences, asking: How do the distribution, duration, and transition of visual attention across functionally organised interface regions relate to gameplay outcomes?

To investigate this research question, this study uses eye tracking to characterise how players allocate, sustain, and shift attention across functionally categorised AOIs, and examines how these gaze behaviours relate to gameplay outcomes. We situate our investigation in a deck-building game with a turn-based combat interface that requires ongoing situation assessment and strategic decisions. Core interface elements remain spatially stable across sessions, enabling comparison across participants. At the same time, the interface combines persistent overlays with evolving combat events, prompting players to shift attention between state monitoring and strategic adjustment. Within this controlled yet engaging context, our study asks:
\begin{itemize}
\item \textbf{RQ1.} How do the distribution and duration of visual attention across functionally organised interface regions relate to gameplay outcomes?
\item \textbf{RQ2.} How do transition probabilities and the structural complexity of gaze transitions between these regions relate to gameplay outcomes?
\end{itemize}

To address these research questions, we conducted a within-subject experiment with 32 participants, recording ocular signals while they engaged in gameplay. Gameplay outcome was defined by the natural result of the game into a Win group and a Loss group. To operationalise gaze behaviour analysis, we organised AOIs into four functional categories based on the structure of the game interface and subsequently refined them using gaze heatmap intensity, yielding six AOI regions. We then extracted five AOI-based measures spanning three event types: AOI hits for spatial distribution, dwell time for sustained attention, and transition-based measures for sequential coordination between AOIs. The transition-based measures included transition probabilities, transition matrix entropy, and stationary entropy, capturing pairwise gaze shifts, the structural complexity of those shifts, and the long-term distribution of attention across AOIs, respectively.
Our results reveal outcome-related differences in gaze allocation and transition patterns across task-relevant interface elements. The Win group exhibited more selective patterns of attention allocation that extended to peripheral resources, whereas the Loss group focused more on enemy information. Regarding gaze shifts, the Win group exhibited more frequent action-oriented gaze transitions and a more dispersed, evenly distributed transition structure, whereas the Loss group demonstrated more frequent recurrent transition patterns.

This study provides empirical evidence that AOI-based gaze behaviour can reveal visual attentional mechanisms associated with gameplay outcomes through gaze allocation and transition patterns. This study contributes to eye tracking and game user research by:
\begin{itemize}
\item Systematically integrating three AOI-based gaze event types within strategy gameplay and identifying outcome-related patterns in how players distribute, sustain, and shift visual attention.
\item Providing a replicable procedure for defining, extracting, and validating functional AOIs in a game interface.
\end{itemize}
\section{Related Work}
Grounded in prior research on visual attention, we organise this section around three perspectives: the cognitive mechanisms of selective attention, including bottom up and top down processes, and how they shape gaze behaviour in interactive contexts; the use of eye tracking in video game research and how gaze behaviour relates to interface characteristics, task demands, and player differences; and AOI based metrics for evaluating gaze distribution, duration, and transition structure.

\subsection{Selective Attention and Gaze Behaviour}
Selective attention enables people to prioritise behaviourally relevant information while filtering competing or irrelevant inputs \citep{desimone1995neural, carrasco2011visual}.
Kahneman’s capacity model provides a classic account of attention as a limited pool of cognitive resources, allocated across competing tasks according to task demands and individual factors of arousal and perceived task importance \citep{kahneman1973attention}. 
Attentional deployment in visual search has been modelled as being guided by both bottom-up stimulus-driven activation and top-down target guidance \citep{wolfe1994guided}.
Bottom-up visual attention is understood as a stimulus-driven process in which low-level visual features, including intensity, colour, and orientation, are combined into a saliency map that prioritises visually conspicuous locations for further processing \citep{itti1998model}. On the other hand, top-down visual attention emphasises the role of current task goals in shaping attentional selection. Early eye movement research showed that gaze patterns vary with viewing instructions even when observers inspect the same image \citep{yarbus1967eye}. Similarly, attentional control accounts propose that competition among visual representations is biased toward information relevant to current goals and behaviour \citep{desimone1995neural, corbetta2002control}.
In visual environments, the interaction of bottom-up sensory and top-down attentional influences creates an integrated saliency map, in which visually strong stimuli and behaviourally relevant information are represented across visual space and guide gaze orientation \citep{treue2003visual}.
Empirical studies in interactive contexts have further shown how these two sources of attentional priority operate during real tasks. 
In virtual reality scenes, saliency based models have simulated visual attention by assigning higher priority to objects within the viewer’s field of view, such as virtual items and avatars, based on properties including proximity, eccentricity, orientation, and velocity \citep{oyekoya2009saliency}. 
Beyond external saliency, \citet{10.1145/3352763} found that task condition affected visual attention in interactive virtual environments, with navigation showing distinct fixation patterns and greater reliance on intensity conspicuity than search or free viewing. 
Within gameplay contexts in which players pursue goals and make decisions under ongoing interaction demands, \citet{Sundstedt2008FixationBehavior} found that saliency maps based on low level visual features had limited predictive power for players' fixation behaviour and emphasised the role of task structure and interaction goals that guide players’ attention. In mobile game navigation interfaces, \citet{jiang2019applying} similarly showed that gaze measures varied across free browsing and task oriented conditions, further indicating that gameplay attention should be interpreted in relation to the task context in which gaze behaviour occurs.

In sum, the reviewed theoretical and empirical work positions selective attention as the prioritisation of visual information under limited cognitive resources. In interactive contexts, these attentional priorities shape gaze patterns, as users orient toward salient visual elements while also adjusting gaze according to task demands. These works provide the theoretical basis for using gaze behaviour to examine how players select and prioritise task relevant information during gameplay.




\subsection{Visual Attention and Performance Related Gaze Behaviour in Video Games}
Eye tracking has long been used in video game research to examine how players visually engage with game environments.
\citet{almeida2011eyes} reviewed visual attention and eye tracking input in video games, framing eye tracking as both an input modality and an evaluation method for examining how players visually interact with game interfaces. The survey discussed prior studies showing that video game play has been associated with changes in visual attention skills, such as attentional capacity, visual search performance, and spatial distribution of attention.
Similarly, \citet{sundstedt2012gazing} provided an introduction to eye tracking in games and virtual environments, with a primary emphasis on gaze based game interaction and control. For diagnostic use, the work summarises studies that examined what players focus on during gameplay, including how fixation behaviour can inform game design, saliency prediction, rendering priorities, and the interpretation of task relevance in games. In game user experience, \citet{zammitto2014gaming} further positioned eye tracking as an emerging evaluation method for game user experience, arguing that gaze behaviour across distinguishable game elements such as menus, overlays, and gameplay cues can inform the evaluation of player experience and support the commercial game design process. Together, these studies establish eye tracking as a well grounded methodology in video game research, providing conceptual foundations and analytical examples for interpreting gaze behaviour in relation to player attention, user experience and game design.

Empirical work has used eye tracking to examine how visual attention is shaped by interface characteristics and task demands during gameplay.
\citet{el2006visual} compared visual attention patterns in an action-adventure game and a first-person shooter game, showing that gaze behaviour differed across game genres and that bottom-up visual cues, such as colour contrast and motion, could attract attention, while top-down goal-directed search tended to dominate players' attention in 3D game environments.
In addition to visual properties of game elements, players’ gaze distribution is also shaped by the spatial organisation of gameplay information. \citet{almeida2016video} combined eye tracking with player movement data, finding that central and strategically significant areas attracted the majority of fixations, whereas peripheral elements were largely ignored. In a simulated racing game, \citet{joyce2024less} evaluated how visual attention is distributed between the primary task area (the track) and peripheral HUD elements. They found that higher-skilled players displayed less spatial allocation to the track and more allocation towards the heads-up display elements.
Beyond general spatial distribution, visual attention is also modulated by specific cognitive demands associated with interface regions. \citet{lan2026impact} examined how reading demand affects attention to the subtitle area during gameplay. By comparing low and high reading demand conditions, they found that increased demand led to higher dwell time percentages, more fixations, and longer fixation durations.
Together, these empirical studies show that localised gaze behaviour during gameplay is sensitive to gameplay context, the spatial organisation of interface information, and both the visual properties and relevance of game elements. They provide important evidence that where players look is shaped by both the visual structure of the game interface and the demands of the ongoing task.

Alongside the influence of game interface characteristics, visual attention also varies across players. Eye tracking studies have shown that differences in gaze behaviour can reveal how players process game information across different levels of experience, knowledge, and task performance.
In game based learning, \citet{conati2013understanding} examined students' attention to adaptive textual hints by defining a single AOI over the hint message. They found that students with the lowest and highest pre test scores made fewer fixations per word on hints than those with intermediate scores. Hint attention also varied with hint timing and attitudes toward help, with greater attention to hints associated with improved subsequent performance.
\citet{lu2021eyes} investigated the relationship between gaze behaviour and spatial navigation abilities in an educational video game. By defining two AOIs, the topographic map and the task area, they analysed average fixations on AOIs, first fixation time, and total visit duration. They found that high performers exhibited focused gaze on the map region, whereas low performers displayed dispersed gaze patterns.
In a static FIFA screenshot viewing task, \citet{wang2024comparative} used expert defined AOIs covering multiple tactical information regions to compare professional and non professional eSports players' visual attention. They analysed fixation-based measures such as fixation counts and fixation duration, and found that professional players showed more tactically oriented and focused visual search. This comparison was based on within-AOI fixation measures over static screenshots viewed passively, leaving open how attention is coordinated during interactive play.
\citet{jeong2022difference} examined gaze control differences between players with high and low skill levels in \textit{StarCraft}. They found that high skill players showed higher fixation ratios in AOIs used to follow task progress and manage unit production. At the whole screen level, high skill players also showed faster and broader visual information acquisition during gameplay, reflected in wider gaze distribution and faster, more frequent saccades.

Overall, the reviewed work shows that gaze behaviour in games is shaped by interface characteristics, task demands, and player related differences. Existing studies have operationalised gameplay information in different ways, including single task relevant AOIs \citep{conati2013understanding, lan2026impact}, broad interface divisions (e.g., central versus peripheral regions, task area versus HUD elements) \citep{almeida2016video, lu2021eyes, joyce2024less}, and multi AOI analyses based on static gameplay screenshots \citep{wang2024comparative}. These studies provide valuable evidence for identifying whether particular game regions attract attention. However, such spatial units have largely been treated independently, with gaze accumulated within each unit and reported variously as fixation counts, fixation durations, fixation percentages, and dwell time. These measures are well suited to understanding where game players are looking by characterising the gaze distribution across predefined regions \citep{sundstedt2012gazing}, but they provide less insight into how gaze is organised between regions as players coordinate spatially distributed gameplay information. 
Video games involve a series of decision processes in which player actions are informed by multiple sources of gameplay information \citep{zammitto2014gaming}. Understanding gaze transitions between interface elements is particularly important when gameplay involves integrating task information that plays different functional roles, such as situation assessment, resource monitoring, and action selection. In this context, transition based analysis provides a way to quantify directional gaze shifts between information sources, revealing which relationships among game elements become more prominent as players coordinate information during decision making. Therefore, a more complete account of player attention needs to integrate multiple dimensions of gaze behaviour: spatial localisation, temporal persistence, and transition structure. These dimensions capture where gaze is directed, how long it is maintained, and how it shifts between AOIs over time.

\subsection{AOI-based Metrics for Evaluating Attention}
AOI based analysis is a common eye tracking approach in which researchers define regions of interest within a visual stimulus or interface, allowing researchers to examine whether gaze is directed to expected areas and what eye movement properties occur within those areas \citep{holmqvist2011eye, duchowski2017eye, blascheck2014state, poole2006eye}. Foundational methodological work in eye tracking has described three core AOI based events: AOI hits, dwells, and transitions \citep{holmqvist2011eye, borys2017eye}. An AOI hit records whether a gaze sample falls within an AOI. Aggregating these hits across viewing time provides a measure of gaze occurrence within each region. Scene viewing research has shown that fixation density varies with the informativeness of scene regions, with more informative regions receiving more fixations \citep{mackworth1967gaze, henderson1999high}. By capturing localised gaze occurrences, hit-based measures allow researchers to map how visual attention is distributed across areas of varying informativeness.
To complement spatial mapping of gaze with temporal information, dwell time measures how long gaze remains within an AOI \citep{holmqvist2011eye, green2002where}. As longer fixations have often been linked to ongoing cognitive processing, dwell based measures provide a way to relate sustained attention to task relevant information \citep{irwin2013fixation, just1976eye, Rayner1998, negi2020fixation}.
However, while AOI hits and dwell time capture localised aspects of visual attention, they do not describe how gaze moves between regions over time.
AOI transitions, also known as ‘gaze shift’, record the movement from one predefined region to another \citep{holmqvist2011eye}. Early work by \citet{ellis1986statistical} formalised this approach by modelling fixation sequences across AOIs as a first-order Markov process, providing the methodological basis for transition matrix analyses of gaze. Grounded in classic scanpath theory \citep{noton1971eye}, sequential eye movements can reflect how observers organise visual features during scene interpretation, connecting distinct visual features in a preferred order to form a coherent mental representation. In the context of interface evaluation, transition-based metrics extend the analysis beyond localised fixations by assessing how observers dynamically coordinate and integrate spatially distributed information \citep{goldberg1999computer}. 
Specifically, systematic gaze shifts between specific regions can serve as a behavioural marker of active comparison between elements or establishment of logical connections between disparate sources of information \citep{russo1983strategies, hegarty1993constructing}.
Moving beyond isolated pairwise connections, gaze transition entropy has been adopted to compare transition matrices and quantify the structural complexity of visual transition patterns \citep{shannon1948mathematical, shiferaw2019gaze, sato2024gaze, shic2008amorphous, jordan2009analysis, krejtz2015gaze}. \citet{krejtz2015gaze} operationalised transition entropy and stationary entropy as AOI based measures for characterising the complexity of gaze switching patterns and the spatial distribution of visual attention across AOIs. Their findings indicate that entropy based gaze measures can capture variation in gaze transition structure associated with viewer related differences such as curiosity and familiarity, while their interpretation remains dependent on the viewing task and stimulus context. Specifically, transition entropy captures the structural unpredictability of gaze shifts between AOIs, reflecting how variable or dispersed the sequence of attentional switches is, whereas stationary entropy captures the long-run distribution of attention across AOIs, reflecting how evenly visual attention is allocated overall \citep{shiferaw2019review}. The two measures therefore index different aspects of gaze organisation and together provide a more complete account of how attention is coordinated across spatially distributed information sources.

In this section, we summarise five AOI-based metrics: AOI hit for fixation occurrence within AOIs, dwell time for gaze duration within AOIs, transition probability for gaze shifts between AOIs, transition entropy for the structural complexity of gaze transitions, and stationary entropy for the overall distribution of gaze across AOIs. Applied to a strategy game interface, these metrics provide a structured way to capture players’ visual attention during gameplay by identifying which areas are visually sampled, how long players remain engaged with them, and how relationships between different information sources are organised through gaze shifts. Grounded in the methodological literature on eye tracking \citep{holmqvist2011eye, krejtz2015gaze}, these established metrics form the analytical basis for examining visual attention in our study. On this basis, we further provide a framework for defining and extracting AOIs from the tested game interface according to their gameplay functions.

\section{Method}
We designed a within subject experiment in a strategy oriented game to investigate how players allocate and shift visual attention during combat, and how these patterns relate to gameplay outcomes. Eye tracking data were recorded during gameplay and segmented into combat episodes. Guided by game mechanics and gaze density maps, we defined AOIs within the combat interface and applied a computational approach to quantify visual attention patterns. Specifically, we computed AOI hit rate, dwell time, transition probability, transition matrix entropy, and stationary entropy to characterise visual attention dynamics across functional AOIs.

\subsection{Participants}
\label{sec:participants}
We recruited 32 participants (16 male, 16 female) from the university campus and online platforms. Participants were aged 18–34 years, with 54.3\% in the 18–24 range and 45.7\% in the 25–34 range. The majority held a bachelor’s degree or higher. To ensure a comparable baseline, only individuals with no prior experience of the game were included. All participants reported normal or corrected-to-normal vision. The study protocol was approved by the university ethics committee, and all participants provided informed consent prior to participation. Participants received either course credit or a voucher as compensation.

\subsection{Experimental Design}
We conducted a within-subject design in which all participants completed the Act~1 portion of a commercial video game, \textit{Slay the Spire} \citep{slaythespire}, to investigate players’ visual attention and strategic decision-making.
The two sessions were consecutive segments of the same run, separated by a predefined progression point. Session 1 began with a tutorial and continued through regular encounters. Session 2 included the final boss fight.

\textit{Slay the Spire} is a turn-based deck-building game built around a core loop of strategic card selection, resource management, and adaptive planning \citep{SlayWiki}. Players act using a customisable deck of cards, where each card represents attacks, defences, or effect actions. During each combat encounter, enemy intent is visible, requiring players to choose cards in response to incoming threats, manage their limited energy and resources, and sequence actions efficiently. Across the run, players progress through successive encounters by building and refining their deck, culminating in a boss fight that concludes the level. The game presents multiple concurrent information sources on the screen, requiring players to continuously shift their gaze between these elements to evaluate threats, plan responses, and select actions, making it well suited for examining how visual attention is distributed across AOIs and how gaze transitions reflect strategic processing. 
The combat interface is visually stable across sessions, with key elements consistently located in the same on screen positions (e.g., the hand cards remaining at the bottom of the screen), supporting systematic AOI mapping and comparability across participants.
In addition, the interface separates persistent status monitoring (energy, health, enemy intent) from turn specific decision information (current hand, card execution, enemy actions), enabling analysis of gaze allocation between state tracking and option evaluation. 

To ensure comparability across participants, we fixed the character and level map using the game’s customisation options and applied a constant run seed, thereby producing an identical layout and enemy sequence across all runs. The entire gameplay was restricted to the first level, corresponding to one fixed map layout. This procedure controlled task content across participants, allowing between-participant comparisons of gaze behaviour under matched encounter sequences.
Each participant began with a short built-in introductory sequence for familiarisation before entering the main combat stages. The first session lasted up to 25 minutes, whereas the second session lasted up to 15 minutes.
Each participant completed one uninterrupted run. Participants who reached and defeated the final boss were coded as Win. All other runs were coded as Loss. Participants were instructed that their goal was to defeat the boss, supporting consistent motivation and engagement throughout the run.

\subsection{Eye Tracking Apparatus}
\begin{figure}
    \centering
    \includegraphics[width=0.8\linewidth]{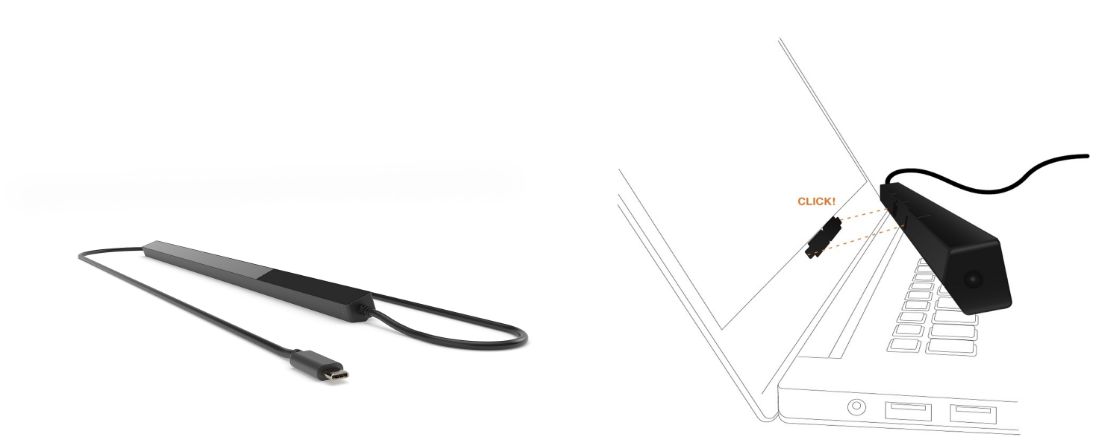}
    \caption{Eye tracking system (Tobii Pro Fusion).}
    \Description[Eye tracker used in the experiment.]
    {The figure shows the Tobii Pro Fusion eye tracker, a rectangular bar-shaped device designed to be positioned below a monitor to record eye movements.}
    \label{fig:eye}
\end{figure}

Eye-tracking data were recorded with the Tobii Pro Fusion, a screen-based binocular eye tracker (see Figure ~\ref{fig:eye}) \citep{TobiiProFusion}. The device sampled at 250 Hz and offered high spatial precision. It was integrated with a 23.8-inch monitor (1920 × 1080 resolution), and participants were seated approximately 60 cm from the screen without a chin rest, allowing for natural head movement. We used Tobii Pro Lab software for calibration and data collection \citep{tobii_accuracy_precision}.
We selected a screen-based tracker to support natural interaction, ensuring stable gaze estimation while maintaining ecological validity for seated digital game play. All recordings were conducted under controlled laboratory conditions, where lighting, sound, and seating arrangements were standardised to minimise environmental variability. 

\subsection{Procedure}  
Participants first completed a demographic questionnaire. At the beginning of the experiment, each participant completed a 9-point calibration and validation procedure in Tobii Pro Lab. Calibration was repeated if the mean accuracy exceeded 1.0° or if data loss was greater than 10\%. All participants met these criteria before proceeding. The first game session began with a built-in tutorial, during which participants could familiarise themselves with the mechanics and ask questions. After the first game session, participants were offered a short break if needed, and the eye tracker was recalibrated to ensure data quality. The second gameplay session, culminating in a boss fight, was then completed.

\subsection{Data Processing}
Raw gameplay recordings and eye-tracking logs were processed to extract combat episodes and derive fixation-level gaze data for subsequent analyses.

\subsubsection{Combat Event Segmentation}  
To analyse gaze behaviours that are highly related to decision making and strategic planning, we first segmented all combat episodes from the complete gameplay recordings.
We used OpenCV’s \citep{opencv_library} image processing techniques to identify distinctive UI elements that consistently appeared during battle sequences. Specifically, multiple interface components that repeatedly appeared across all combat scenes were used for automated retrieval of combat-related video frames, such as the End Turn button, health bars, and the resources panel. Multi-template matching based on normalised cross-correlation (threshold = 0.7) was applied to each frame, ensuring that only frames containing all key combat-related UI elements were retained. The automatically extracted battle segments were manually inspected by the researcher to remove any falsely detected or incomplete sequences. 
The validated frame ranges were then used to derive precise start and end timestamps for each battle episode. These timestamps were aligned with the raw eye-tracking logs, and each gaze sample was assigned to its corresponding battle interval based on its recorded time.

\subsubsection{Gaze Preprocessing and Fixation Detection}
Raw gaze data were cleaned to remove signal loss and invalid samples before fixation identification. Periods of signal loss (e.g., blinks), which accounted for less than 5\% of the dataset, were corrected using linear interpolation to maintain temporal continuity. Data points located outside the display boundaries were removed to ensure that only valid on-screen gaze behaviour was analysed. Fixation events were identified using Tobii’s default I-VT algorithm. Fixations shorter than 60\,ms were excluded \citep{holmqvist2011eye}, and additional boundaries were introduced when abnormal timestamp gaps occurred, defined as greater than four times the median sampling interval. Each fixation was represented by its centroid (mean normalised gaze coordinates) and its duration, which were used for subsequent AOI-based analyses.

\subsection{AOI Definition and Extraction}  
\label{sec:aoi}
\begin{figure}[t]
    \centering
    \captionof{table}{Candidate AOIs in \textit{Slay the Spire} combat, grouped into four functional categories}
    \label{tab:game_elements}

    \renewcommand{\arraystretch}{1.2}
    \begin{tabularx}{\linewidth}{
        @{}
        >{\raggedright\arraybackslash}p{3cm}
        >{\raggedright\arraybackslash}p{3cm}
        >{\raggedright\arraybackslash}X
        @{}
    }
        \toprule
        \textbf{Category} & \textbf{Game Element} & \textbf{Description} \\
        \midrule
        Enemy Information & Enemy Intent, Enemy Status &
        Indicates the opponent’s upcoming action and current combat state, guiding whether the player should attack, defend, or adapt strategy. \\
        \midrule
        Player Information & Player Status &
        Displays the player’s current combat condition, informing survivability evaluation and tactical decision-making. \\
        \midrule
        Action Resources & Hand Cards, Energy, Draw Pile &
        Indicates available actions, resource constraints, and upcoming card availability. \\
        \midrule
        Auxiliary Resources & Potions, Relics &
        Provide temporary bonuses or passive modifiers that support survival and shape long-term strategy beyond a single battle. \\
        \bottomrule
    \end{tabularx}
    \vspace{1.2em}

    \captionsetup{type=figure}
    \includegraphics[width=\linewidth]{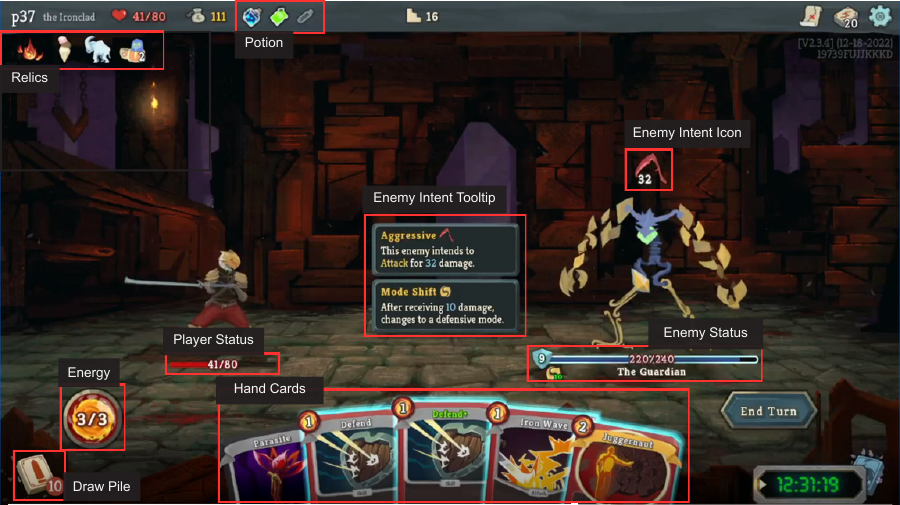}
    \caption{Top-down candidate AOIs identified from the \textit{Slay the Spire} combat interface.
    Highlighted regions indicate representative interface elements used as initial AOI candidates.}
    \label{fig:battle_scene}
    \Description[Combat interface with candidate AOIs highlighted]{
A top-down view of the Slay the Spire combat interface with candidate areas of interest highlighted. Rectangular overlays mark representative interface elements selected as initial AOI candidates, including enemy-related information, player status, energy indicators, hand cards, and auxiliary interface panels. The highlighted regions illustrate how functional interface elements are spatially delineated prior to AOI boundary localisation.
}
\end{figure}
\begin{figure}
    \centering
    \includegraphics[width=\linewidth]{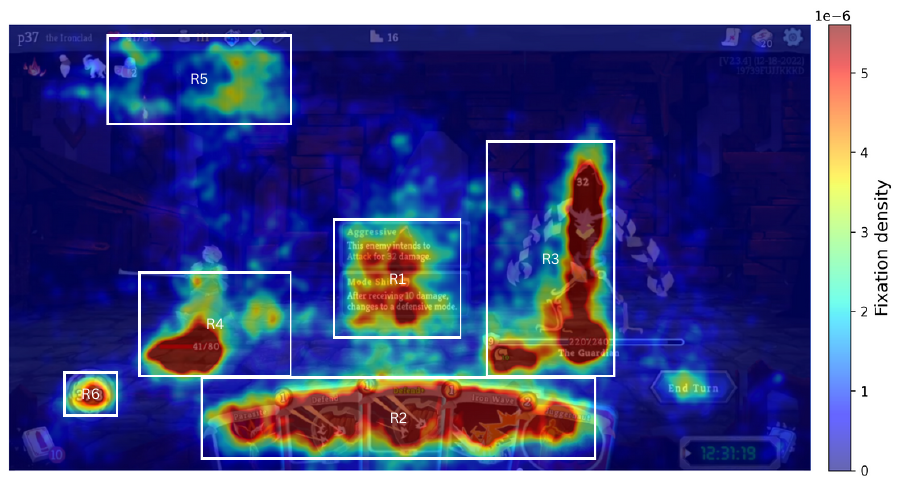}
    \caption{Fixation-based gaze density map during the combat, overlaid with the final AOIs (R1--R6) identified through bottom-up refinement. 
Colours indicate normalised fixation density, with warmer colours representing higher visual attention. 
White bounding boxes denote the spatial extents of the final AOIs derived from the gaze density distribution.}
    \label{fig:density}
    \Description[Fixation density map with final AOI boundaries]{
A fixation-based gaze density map overlaid on the \textit{Slay the Spire} combat interface. Colour shading represents normalised fixation density, with warmer colours indicating higher fixation concentration. White rectangular bounding boxes labelled R1 to R6 mark the spatial extents of the final areas of interest, which were determined through bottom-up refinement based on the gaze density distribution.
}
\end{figure}
\begin{figure}
    \centering
    \includegraphics[width=\linewidth]{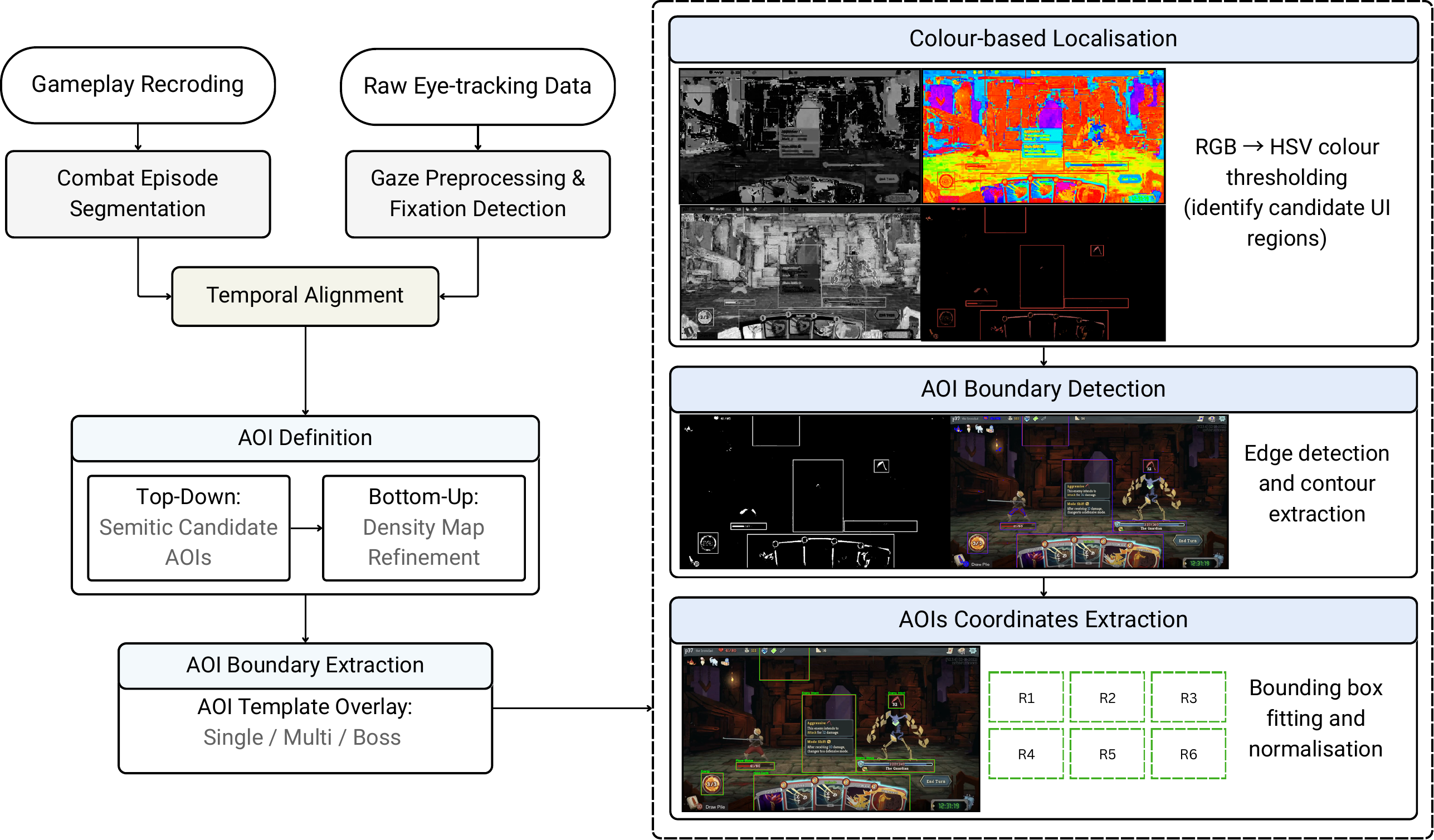}
    \caption{Overall pipeline for gameplay–eye tracking temporal alignment, AOI definition, boundary localisation, and extraction.}
    \label{fig:aoi_pipeline}
    \Description[Pipeline diagram for AOI definition and extraction]{
A flow diagram illustrating the eye-tracking analysis pipeline during gameplay. The diagram is composed of sequential connected boxes showing: gameplay recording and eye-tracking data acquisition; temporal alignment between gaze data and gameplay phases; definition of functional areas of interest on the game interface; localisation of AOI boundaries across different gameplay stages including early, mid, multi-enemy, and boss encounters; and extraction of AOI-based gaze measures for subsequent analysis.
}
\end{figure}

To analyse gaze behaviour in combat, we adopted a hybrid top-down and bottom-up approach to define AOIs.
This process consisted of three stages: identifying candidate AOIs based on interface functionality, refining these candidates using gaze density patterns, and subsequently localising the final AOIs as spatial regions on the game interface. 
An overview of the AOI definition and extraction workflow is provided in Figure~\ref{fig:aoi_pipeline}.

\subsubsection{Top-down Candidate AOIs}
In the top-down stage, candidate AOIs were defined based on the functional components of the game interface, following the in-game mechanics described in the official \textit{Slay the Spire} documentation \citep{SlayWiki}.
Each AOI corresponds to a distinct source of visual and strategic information, grouped into four functional categories: 
\textit{Enemy Information}, \textit{Player Information}, \textit{Action Resources}, and \textit{Auxiliary Resources}. 
Table~\ref{tab:game_elements} summarises these AOIs along with their in-game functions and classification rationale, and Figure~\ref{fig:battle_scene} illustrates their spatial layout on the combat interface.
These candidate AOIs were subsequently refined using fixation-based gaze density maps to identify regions of concentrated visual attention, while the functional categorisation derived in this stage was retained throughout the analysis to interpret AOI-level findings in relation to gameplay roles.

\subsubsection{Bottom-up Gaze Density Map}
To empirically refine the top-down candidate AOIs, fixation-based gaze density maps were computed across participants. Figure~\ref{fig:density} presents fixation-based gaze-time density maps during the boss combat for all participants. Each fixation was weighted by its duration and normalised to total viewing time, yielding relative gaze time per pixel. The boss combat phase was selected for density-based refinement as it represents the most attention-demanding combat condition, providing the clearest distribution of visual attention across interface elements. The colour bar represents relative gaze-time density, with blue indicating the lowest density and red indicating the highest, scaled to the 99th percentile of all pixel values to reduce the impact of extreme outliers.
To extract regions of elevated visual attention, contiguous areas of elevated density were identified from the resulting heatmap. Regions were extracted from the normalised gaze density map using percentile-based thresholding, with density thresholds applied at levels of at least the 70th percentile. For each threshold, connected components were detected after morphological closing using an elliptical kernel of size 11. Regions with an area smaller than 80 pixels or an integrated density mass below 0.1\% were discarded. To avoid redundant detections across threshold levels, regions whose centroids were within 120 pixels were merged. Threshold values for density percentile and minimum region area were selected based on empirical inspection of fixation density distributions, with candidate elements that did not exceed these thresholds excluded. 
The final set of regions was obtained by ranking candidates by attention mass; the top six regions were retained, corresponding to the candidate AOIs identified in the top-down stage that received concentrated visual attention (excluding sparsely attended elements: the Draw Pile and Relics). 
These regions corresponded to interface components associated with high-intensity visual attention: R1: enemy intent area; R2: player hand (card selection area); R3: enemy status; R4: player status; R5: potion panel; and R6: energy indicator.

\subsubsection{AOI Boundary Extraction}
For the six AOIs (Figure~\ref{fig:density} R1--R6), spatial boundaries were subsequently localised on the game interface to support gaze-based feature extraction.
For non-diegetic UI elements, including Hand Cards (R2), Player Status (R4), Potion Panel (R5), and Energy (R6), AOI frame positions remained relatively stable across battle rounds. In contrast, enemy-related interface elements, including Enemy Intent (R1) and Enemy Status (R3), exhibited positional variation driven by differences in enemy number and layout.
Based on these structural differences, combat encounters were categorised into three battle types (Boss, Single Enemy, Multiple Enemies), with separate AOI boundary templates defined for each type. Example AOI templates illustrating the three battle types are provided in Appendix~\ref{app:boundary} for reference.
To obtain AOI coordinates, we employed an image-processing-based procedure leveraging colour-based segmentation and contour detection applied to manually overlaid AOI boundaries on representative combat screenshots \citep{opencv_library}. 
Recent foundation segmentation models, such as Segment Anything and SAM 2, have provided powerful alternatives for visual object detection \citep{kirillov2023segment, ravi2024sam2}. However, AOI localisation in our study concerned functionally interpretable regions of a structured game interface rather than object segmentation or frame by frame tracking of dynamic game elements. 
Most of these regions occupied relatively stable positions across battle rounds, while enemy related regions varied within predictable layout categories determined by enemy number and arrangement. Given these interface characteristics, the required output was a set of inspectable coordinate boundaries that could be normalised and applied consistently for gaze mapping. A classical image processing pipeline therefore provided a more direct and lightweight solution, as it could isolate visually marked interface regions through explicit colour, edge, and contour based criteria without introducing additional model dependent decisions such as prompt specification, inference variability, or post hoc mask selection.

Specifically, colour segmentation, Canny edge detection, and contour based processing apply explicit image based criteria to recover boundary geometry, allowing the same representative layout to produce the same AOI coordinates under the same processing parameters \citep{canny1986computational, opencv_library}. This procedure also avoided additional model dependent steps, such as prompt selection, inference variability, and post hoc mask selection, which would have introduced unnecessary complexity for a fixed game interface layout. The resulting pipeline therefore provided a lightweight, auditable, and repeatable way to convert the six density-derived AOI regions into normalised screen coordinates across the three battle layout types.

The screenshots were first converted from RGB to HSV colour space, and segmented using predefined colour ranges corresponding to target AOI components. Edge information was extracted using Canny edge detection, and contours were identified to capture the visual boundaries of the target UI elements.
Bounding rectangles were then fitted to these contours to record the spatial extents of each predefined AOI. AOI boundaries were extracted once per battle type and reused across encounters of the same type, balancing computational efficiency and localisation accuracy. All coordinates were subsequently normalised by screen resolution and applied consistently for gaze mapping within each battle type.

To assess the stability of template-based AOI extraction across encounters, we performed visual inspection across multiple battle videos. Boss and Single-enemy templates were each validated against five participants' battle videos, while Multi-enemy templates were validated against nine participants' videos to capture greater variability in enemy combinations. Within each battle type, 15 frames were randomly sampled from each selected battle video using uniform frame-index sampling across the central 80\% of each video duration. After excluding noisy frames that did not contain combat content (e.g., transition or animation frames), we extracted 156 validation frames in total. For each sampled frame, we overlaid the corresponding battle-type AOI template onto the frame using the normalised AOI coordinates. We visually inspected whether each of the six AOI boundaries encompassed the corresponding target UI element, and recorded the number of aligned AOIs per frame. Across frames, the proportion of target UI elements falling within their AOI boundaries was 95.4\%, with reduced alignment observed primarily for the Enemy Intent AOI. Detailed alignment results across battle types and AOIs are reported in Appendix~\ref{app:aoi_validation}.

\subsection{AOI-Based Gaze Measures}  
Five gaze-based measures were computed across three AOI event types (Table~\ref{tab:aoi_events}): AOI Hit and AOI Dwell Time at the within-AOI level, and AOI Transition Probability, transition matrix entropy ($H_t$), and stationary entropy ($H_s$) at the between-AOI level. Each measure was computed per participant and normalised by a baseline: gaze sample counts for AOI Hit, total phase duration for Dwell Time, outgoing transition counts for AOI Transition Probability, and the theoretical maximum entropy for $H_t$ and $H_s$, ensuring comparability across sessions.
\begin{table}[t]
    \centering
    \caption{AOI-based gaze metrics for trial-level analysis. Each metric captures a distinct aspect of attentional behaviour.}
    \label{tab:aoi_events}
    \resizebox{\textwidth}{!}{%
    \begin{tabular}{>{\raggedright\arraybackslash}p{2.8cm} >{\raggedright\arraybackslash}p{7.2cm} >{\raggedright\arraybackslash}p{6.5cm}}
        \toprule
        \textbf{Metric} & \textbf{Description} & \textbf{Criteria} \\
        \midrule
        \textbf{AOI Hit (AH)} & Proportion of gaze samples that fell within a specific AOI, indicating allocation of attention across interface elements. & Screen coordinates normalised to [0,1]; each gaze sample assigned to at most one AOI (priority rule applied for overlaps). \\
        \midrule
        \textbf{Dwell Time (DT)} & Total time spent within an AOI, reflecting sustained engagement with specific interface elements. & Consecutive gaze samples aggregated; minimum fixation threshold of 60\,ms applied. \\
        \midrule
        \textbf{AOI Transition Probability (TP)} & Frequency of gaze shifts between AOIs, reflecting dynamic exploration and attentional switching. & Transition defined as successive fixations between AOIs exceeding 0.5$^\circ$ of visual angle. \\
        \midrule
\textbf{Transition Matrix Entropy ($H_t$)} & Uncertainty of AOI switching, indexing the structural complexity of attentional allocation. & Computed from the per-participant AOI transition probability matrix using Shannon entropy~\cite{holmqvist2011eye}; normalised by maximum entropy. \\
\midrule
\textbf{Stationary Entropy ($H_s$)} & Uniformity of long-term attentional distribution across AOIs, derived from the steady-state of the transition matrix. & Computed from the stationary distribution of the per-participant transition matrix via eigenvalue decomposition~\cite{krejtz2015gaze}; normalised by $\log_2(N)$. \\
        \bottomrule
    \end{tabular}}
\end{table}

AOI Hit (AH) was calculated on a sample basis as the proportion of gaze points that fell within a given AOI \cite{holmqvist2011eye}. Screen coordinates were normalised to the $[0,1]$ range, and each gaze sample was checked against AOI boundaries derived from template-matched UI positions. In cases where AOIs overlapped, a priority rule was applied (e.g., \textit{Enemy Intent} > \textit{Enemy Status} > others) to ensure that each gaze sample was assigned to at most one AOI. For participant $i$ and AOI $j$,  
\begin{equation}
AH_{ij} = \frac{\text{Gaze Points in AOI}_j}{\text{Total Gaze Points}_i} \times 100
\end{equation}

Dwell Time represented the proportion of time spent fixating within an AOI relative to the total duration of the gameplay phase \cite{holmqvist2011eye}. Consecutive gaze samples located within the same AOI were aggregated into fixation events, with a minimum fixation threshold of 60\,ms applied to exclude spurious samples. For participant $i$ and AOI $j$, dwell time was computed as:
\begin{equation}
DT_{ij} = \frac{\sum \text{Fixation Duration in AOI}_j}{\text{Total Time in Phase}_i}
\end{equation}

AOI transition probability measured the relative frequency of directed gaze shifts between pairs of AOIs \cite{holmqvist2011eye}. A transition was defined as successive fixations landing in two distinct AOIs, separated by at least 0.5$^\circ$ of visual angle. For each participant $i$, the probability of a directed transition from AOI $A$ to AOI $B$ was defined as the proportion of that transition type among all observed AOI-to-AOI transitions:
\begin{equation}
TP_i(A \to B) =
\frac{n_i(A \to B)}
{\sum_{X}\sum_{Y \ne X} n_i(X \to Y)} ,
\end{equation}
where $n_i(A \to B)$ denotes the number of observed transitions from $A$ to $B$ for participant $i$.

Transition matrix entropy ($H_t$) quantified the uncertainty of a gaze transition matrix by measuring how transitions were distributed across its cells \cite{holmqvist2011eye, shannon1948mathematical}. For each participant, $H_t$ was computed from the aggregated AOI transition probability distribution derived across gameplay. $H_t$ for participant $i$ and its normalised form were defined as:
\begin{equation}
H_i = - \sum_{k=1}^{K} p_{ik} \log_2(p_{ik}), \qquad H_{\mathrm{norm},i} = \frac{H_i}{\log_2(K)},
\end{equation}
where $p_{ik}$ denotes the probability of a directed transition type $k$, with each transition corresponding to a shift from one AOI to another, and $K = N(N-1)$ denotes the total number of possible directed transitions among $N$ AOIs. Normalisation by the theoretical maximum entropy ensures comparability across participants with different AOI coverage and transition sparsity, isolating differences in transition structure rather than the number of available transitions. Entropy values were then averaged across participants within each performance group to obtain group-level estimates of transition variability.

We further computed the stationary entropy ($H_s$) following Krejtz et al.~\cite{krejtz2015gaze} to quantify how visual attention is distributed across AOIs in the long run. For each participant, the row-normalised transition matrix $\mathbf{P}$ was constructed from observed AOI transitions, with rows corresponding to source AOIs having no outgoing transitions filled with a uniform distribution $1/N$ to ensure regularity~\cite{krejtz2015gaze}. The stationary distribution $\boldsymbol{\pi}$, satisfying $\boldsymbol{\pi}\mathbf{P} = \boldsymbol{\pi}$, was obtained via eigenvalue decomposition of $\mathbf{P}$ and represents the long-term proportion of fixations at each AOI under the fitted Markov chain. Stationary entropy and its normalised form were defined as:
\begin{equation}
H_{s,i} = -\sum_{j=1}^{N} \pi_{ij} \log_2(\pi_{ij}),
\qquad
H_{s,i}^{\text{norm}} = \frac{H_{s,i}}{\log_2(N)}.
\end{equation}

\section{Results}
This section presents the analyses of AOI hit rate, AOI dwell time, and AOI transition patterns between the Win and Loss groups. AOI hit rate and dwell time were compared using the Mann--Whitney U (MWU) test. For AOI transitions, we report transition probabilities visualised as gaze transition networks and a difference heatmap, followed by transition matrix entropy and stationary entropy results quantifying the structural complexity and long-term distribution of gaze shifts across AOIs.

\subsection{AOI Hit Rates Between Win and Loss Groups}
We first calculated the AOI hit rate. To account for differences in overall gameplay duration, hit counts were normalised by dividing the absolute number of AOI hits by the total number of gaze samples per participant. Group differences were assessed using MWU tests, with $p$ values adjusted for multiple comparisons using the Benjamini–Hochberg false discovery rate procedure. Table~\ref{tab:aoi_hit} presents the descriptive statistics of AOI hit rates for the win and loss groups.
\begin{table}[htbp]
  \centering
  \caption{MWU test results for AOI hit rates (\%) between Win group ($n = 16$) and Loss group ($n = 16$) across the whole game. Raw $p$ values and FDR corrected $p$ values (Benjamini and Hochberg procedure) are reported. Effect size $r$ is the rank biserial correlation computed from the Mann Whitney $U$ statistic.}
  \label{tab:aoi_hit}
  \setlength{\tabcolsep}{5pt}
  \renewcommand{\arraystretch}{1.2}
  \begin{adjustbox}{max width=\linewidth}
  \begin{tabular}{
    @{}
    l
    l
    c
    c
    c
    c
    c
    c
    @{}
  }
    \toprule
    Category & AOI & Win M (SD) & Loss M (SD) & $U$ & $p$ & $p_{\mathrm{FDR}}$ & $r$ \\
    \midrule
    Enemy Information
      & Enemy Intent
      & 23.7 (4.1)
      & 26.0 (3.0)
      & 89.0
      & 0.15
      & 0.29
      & 0.30 \\
      & Enemy Status
      & 6.7 (1.4)
      & 7.0 (1.5)
      & 117.0
      & 0.69
      & 0.69
      & 0.09 \\
    \midrule
    Action Resources
      & Hand Cards
      & 32.5 (5.6)
      & 30.8 (3.1)
      & 157.0
      & 0.28
      & 0.42
      & 0.23 \\
      & Energy
      & 1.3 (0.3)
      & 1.6 (0.7)
      & 87.0
      & 0.13
      & 0.29
      & 0.32 \\
    \midrule
    Player Information
      & Player Status
      & 3.2 (1.0)
      & 3.6 (1.3)
      & 104.0
      & 0.38
      & 0.45
      & 0.19 \\
    \midrule
    Auxiliary Resources
      & Potions
      & 3.0 (1.4)
      & 1.7 (1.6)
      & 197.0
      & \textbf{0.01}
      & \textbf{0.06}
      & 0.54 \\
    \bottomrule
  \end{tabular}
  \end{adjustbox}
\end{table}
Across categories, the \textit{Action Resources} category received the highest visual attention, with the majority of fixations directed to \textit{Hand Cards} (Win: 32.5\%, Loss: 30.8\%). Substantial fixation was also allocated to \textit{Enemy Intent} (Win: 23.7\%, Loss: 26.0\%).
Across outcome groups, the win group showed higher hit rates on \textit{Hand Cards} (Win: 32.5\% vs.\ Loss: 30.8\%) and \textit{Auxiliary Resources} (Win: 3.0\% vs.\ Loss: 1.7\%), whereas the loss group allocated relatively more attention to \textit{Enemy Intent} (Loss: 26.0\% vs.\ Win: 23.7\%), \textit{Energy} (Loss: 1.6\% vs.\ Win: 1.3\%), and \textit{Player Status} (Loss: 3.6\% vs.\ Win: 3.2\%). None of these differences reached statistical significance.
The largest between group contrast emerged for \textit{Auxiliary Resources}. The difference was significant at the uncorrected level ($p = 0.010$) but did not remain significant after FDR correction ($p_{\mathrm{FDR}} = 0.059$). Nevertheless, the effect size was large ($r = 0.54$), with substantial distributional separation between groups.

\subsection{AOI Dwell Time Between Win and Loss Groups}
\begin{figure}
    \centering

    \begin{minipage}[t]{0.665\linewidth}
        \centering
        \includegraphics[width=\linewidth]{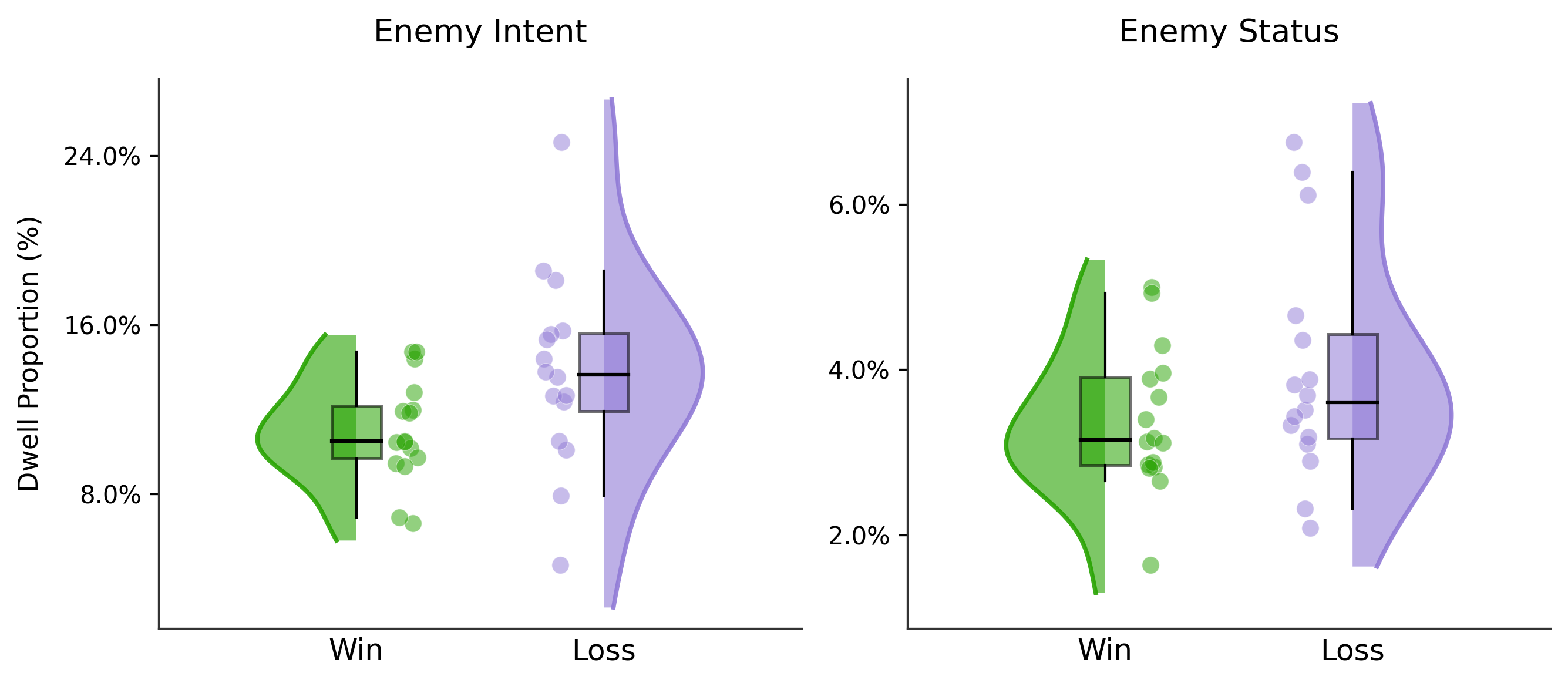}
    \end{minipage}%
    \begin{minipage}[t]{0.335\linewidth}
        \centering
        \includegraphics[width=\linewidth]{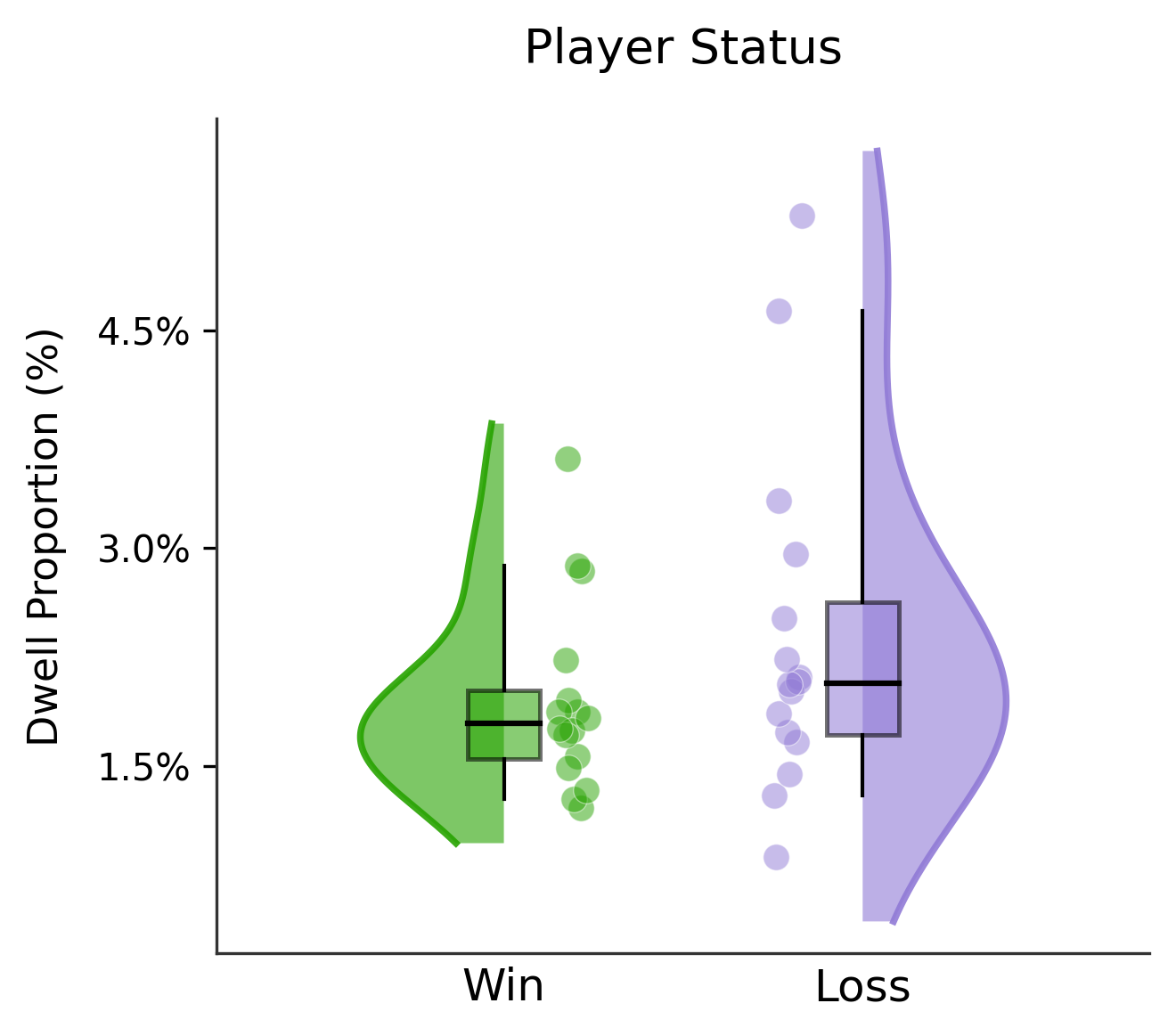}
    \end{minipage}

    \vspace{0.6em}

    \begin{minipage}[t]{0.665\linewidth}
        \centering
        \includegraphics[width=\linewidth]{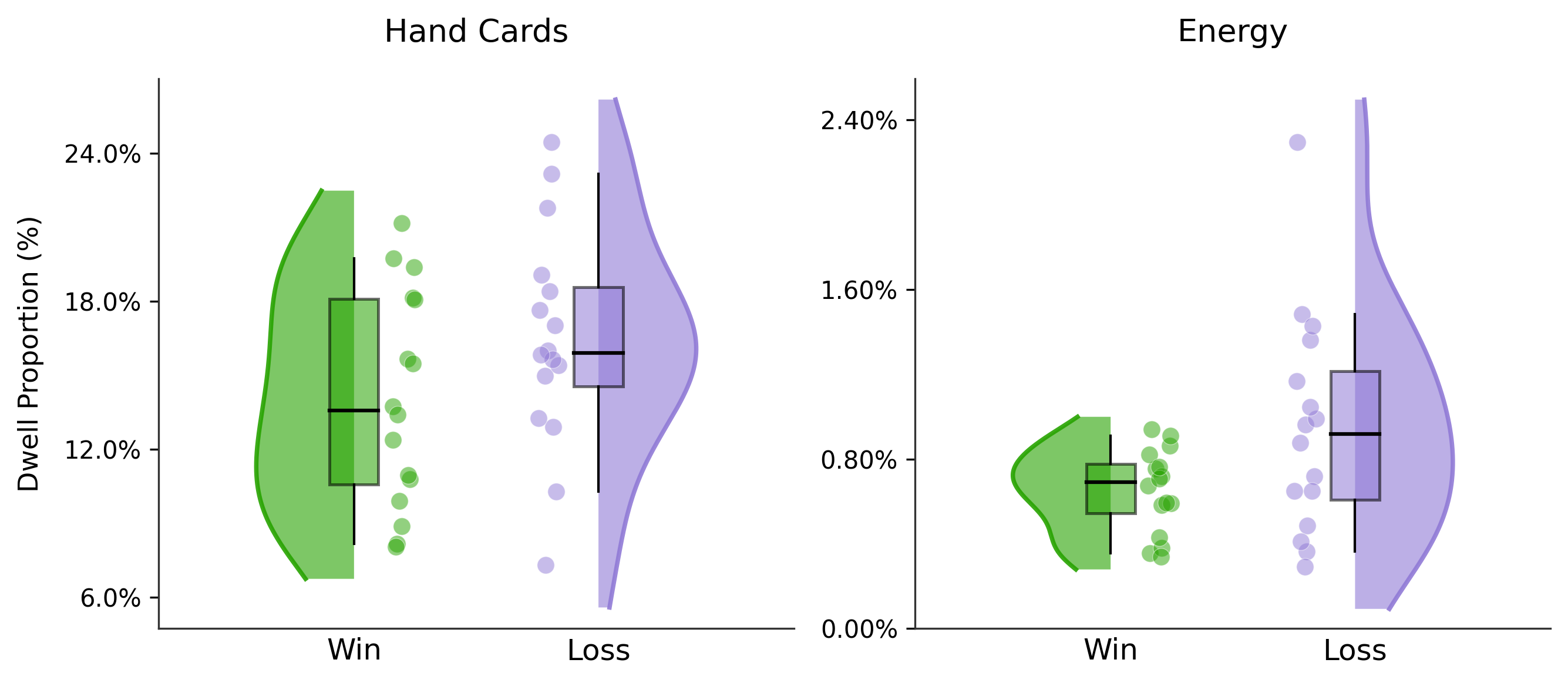}
    \end{minipage}%
    \begin{minipage}[t]{0.335\linewidth}
        \centering
        \includegraphics[width=\linewidth]{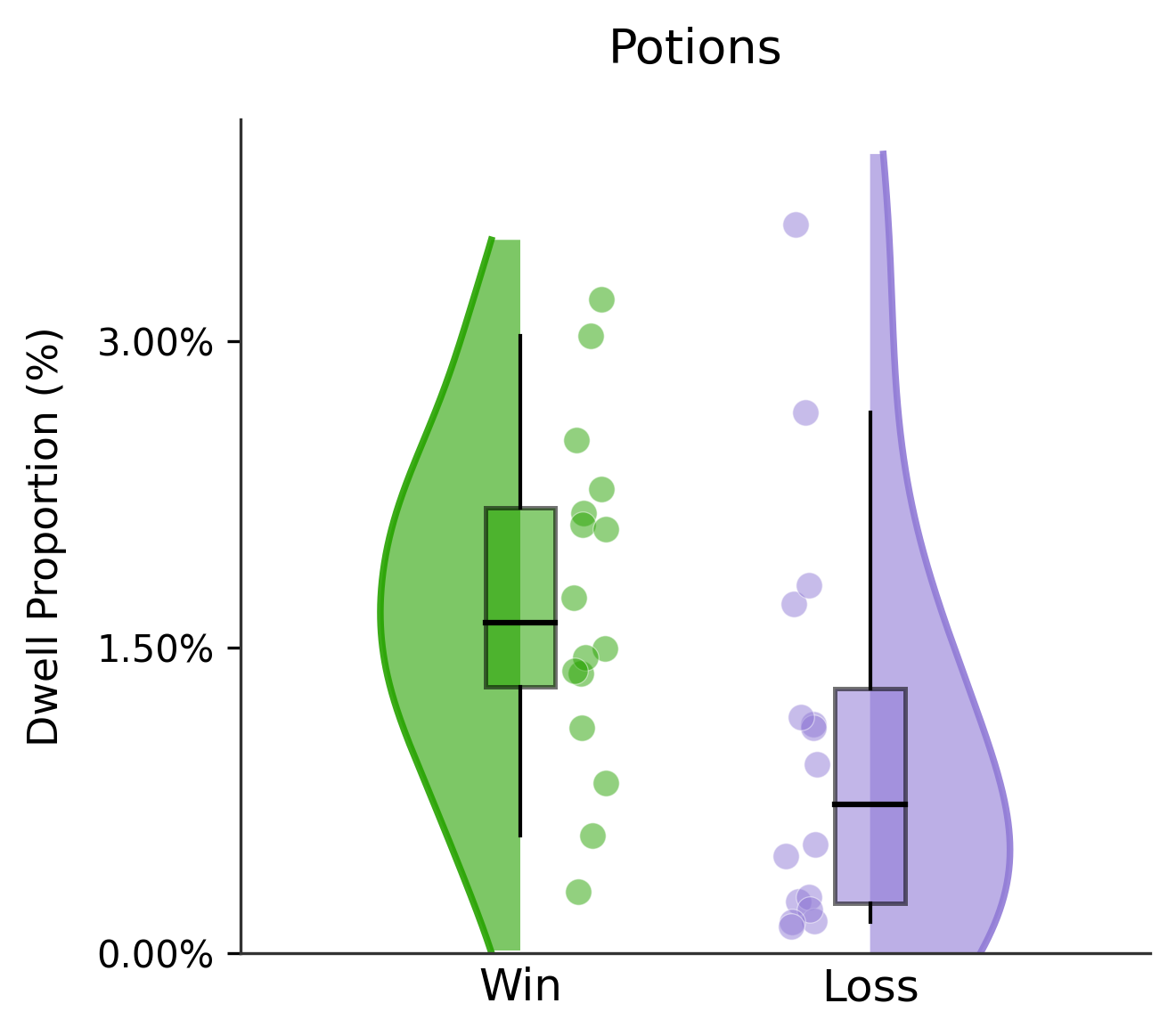}
    \end{minipage}
       \captionof{figure}{Distribution of dwell-time proportions across AOIs for Win and Loss players. Raincloud plots depict participant-level distributions, with boxplots showing the median and interquartile range. The AOIs comprise Enemy Intent, Enemy Status, Player Status, Hand Cards, Energy, and Potions.}
        \label{fig:dwell_aoi}
    \vspace{1.2em}

    \begin{minipage}{\linewidth}
        \centering
        \captionof{table}{MWU test results for AOI dwell time between Win ($n = 16$) and Loss ($n = 16$) groups. Raw $p$ values and FDR-corrected $p$ values are reported. Effect size $r$ denotes the rank-biserial correlation derived from the Mann--Whitney $U$ statistic.}
        \label{tab:aoi_dwell}

        \vspace{0.4em}

                \setlength{\tabcolsep}{5pt}
        \renewcommand{\arraystretch}{1.2}
        
        \begin{adjustbox}{max width=\linewidth}
        \begin{tabular}{l l c c c c c c}
            \toprule
            Category & AOI & Win M (SD) & Loss M (SD) & $U$ & $p$ & $p_{\mathrm{FDR}}$ & $r$ \\
            \midrule
            Enemy Information & Enemy Intent & 11.0 (2.4) & 13.8 (4.6) & 70.0 & \textbf{0.03} & 0.09 & 0.45 \\
            & Enemy Status & 3.4 (0.9) & 4.0 (1.4) & 95.0 & 0.22 & 0.22 & 0.26 \\
            \midrule
            Action Resources & Hand Cards & 14.0 (4.4) & 16.4 (4.5) & 92.0 & 0.18 & 0.22 & 0.28 \\
            & Energy & 0.7 (0.2) & 0.9 (0.5) & 81.0 & 0.08 & 0.16 & 0.37 \\
            \midrule
            Player Information & Player Status & 1.9 (0.7) & 2.4 (1.2) & 95.0 & 0.22 & 0.22 & 0.26 \\
            \midrule
            Auxiliary Resources & Potions & 1.7 (0.8) & 1.0 (1.0) & 192.0 & \textbf{0.02} & 0.09 & 0.50 \\
            \bottomrule
        \end{tabular}%
        \end{adjustbox}
    \end{minipage}

    \Description[Dwell-time distributions across AOI categories for Win and Loss players]{
A composite figure showing dwell-time proportions across AOIs for Win and Loss groups. The figure contains six violin plots arranged in two rows. The top-left panel shows AOIs related to enemy information, and the top-right panel shows AOIs related to player information. The bottom-left panel shows AOIs related to action resources, and the bottom-right panel shows AOIs related to auxiliary resources. In each panel, violin plots depict participant-level distributions of dwell-time proportions, with overlaid boxplots indicating the median and interquartile range for each group. Below the plots, a table reports Mann--Whitney U test results comparing Win and Loss groups for each AOI, including group means and standard deviations, U statistics, raw p values, FDR-corrected p values, and rank-biserial effect sizes.
}
\end{figure}

Differences in AOI dwell time between win and loss groups were assessed for each AOI using the MWU test. To account for multiple comparisons across AOIs, $p$ values were adjusted using the Benjamini--Hochberg FDR correction. Figure~\ref{fig:dwell_aoi} presents the raincloud distributions of AOI dwell time for each performance group, while Table~\ref{tab:aoi_dwell} reports the corresponding statistical results. 
Within the \textit{Enemy Information} category, loss players exhibited broader distributions and higher median dwell times for both \textit{Enemy Intent} and \textit{Enemy Status}. The group difference for \textit{Enemy Intent} was significant at the uncorrected level (raw $p = .03$), accompanied by a large effect size ($r = .45$), but did not remain significant after FDR correction ($p_{\mathrm{FDR}} = .09$).
For the \textit{Player Information} category, represented by the \textit{Player Status} AOI, loss players exhibited a distribution with an upper tail extending towards higher dwell times, whereas the win group showed a more concentrated distribution.
Within the \textit{Action Resources} category, \textit{Hand Cards} accounted for a substantial proportion of visual attention, with values reaching approximately 25\%. The loss group exhibited a higher median dwell time and greater variability, whereas the win group showed a broader distribution centred at lower values. Both groups devoted comparatively limited dwell time to \textit{Energy}, with median values remaining below 1\%. The loss group again displayed a more dispersed distribution extending towards higher values, whereas the win group exhibited a lower and more concentrated distribution.
For the \textit{Potions} AOI within the \textit{Auxiliary Resources} category, the win group exhibited a distribution centred at higher dwell values. In contrast, most observations in the loss group were concentrated below 1.5\%, with a pronounced upper tail. The group difference was significant at the uncorrected level (raw $p = .02$), accompanied by a large effect size ($r = .50$), but did not remain significant after FDR correction ($p_{\mathrm{FDR}} = .09$).

Overall, dwell time was unevenly distributed across AOIs. \textit{Enemy Intent} and \textit{Hand Cards} accounted for the largest proportions of visual attention, whereas dwell time for \textit{Energy} and \textit{Potions} remained comparatively low. Across performance groups, dwell time tended to be higher in the loss group for most AOIs, with \textit{Potions} as the exception. Uncorrected group differences were observed for \textit{Enemy Intent} and \textit{Potions}, but neither remained significant after FDR correction.

\subsection{AOI Transition Analysis Between Win and Loss Groups}
We conducted three analyses to compare AOI transition patterns between the Win and Loss groups. First, transition probabilities were derived and visualised as directed gaze transition networks and a heatmap to illustrate group differences. Second, transition matrix entropy was calculated to quantify the uncertainty of gaze shifts across directed AOI pairs. Third, stationary entropy was computed to characterise the long-term distribution of attention across AOIs.

\subsubsection{Transition Probability Patterns }
\begin{figure}[h]
    \centering
    \includegraphics[width=0.95\linewidth]{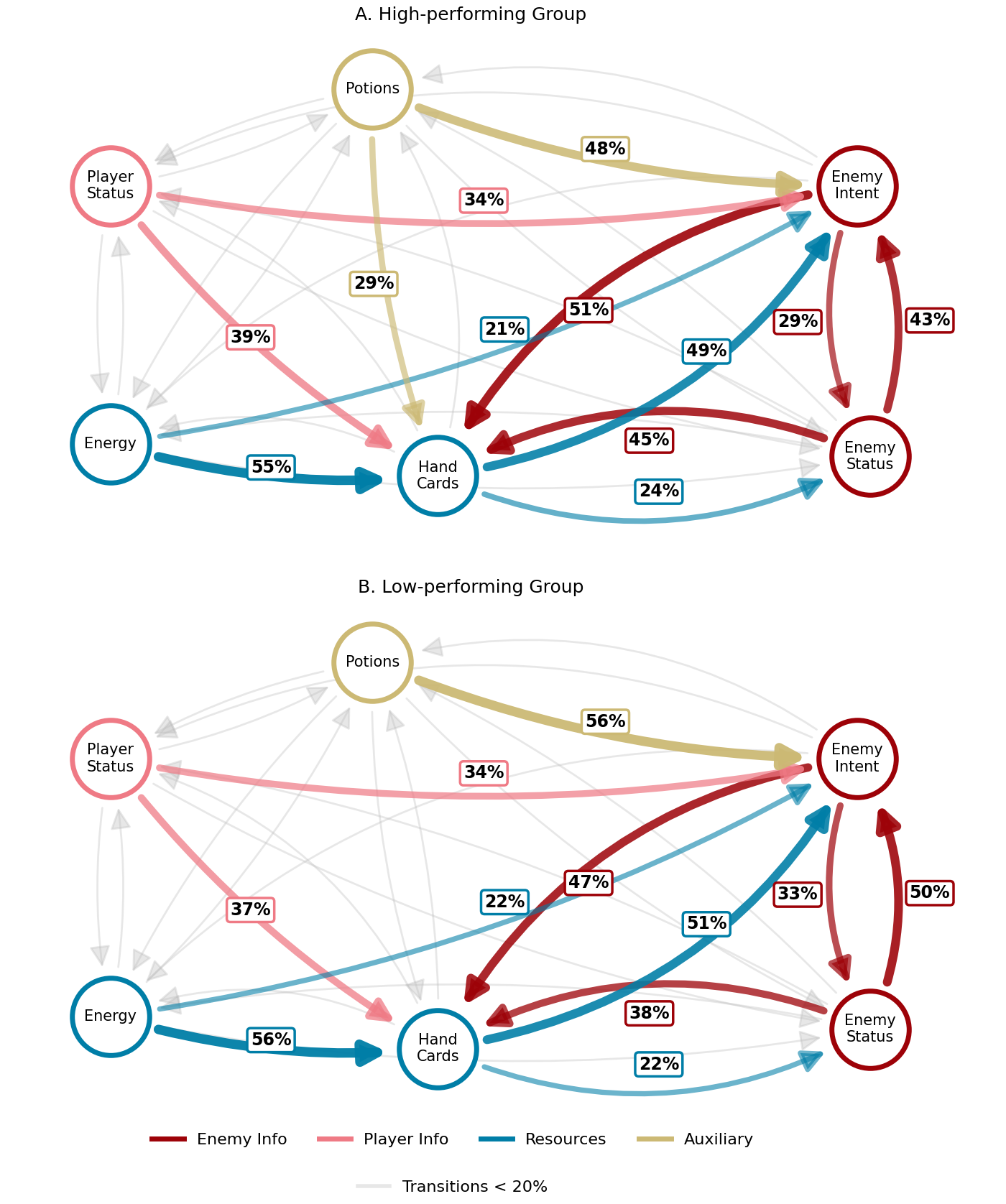}
    \caption{Gaze transition networks between interface elements.
Nodes represent functional AOIs, and directed edges indicate the probability of gaze transitions between AOIs. Edge thickness corresponds to transition probability, and colours denote AOI categories.
Panels show results for the Win group (A) and the Loss group (B), respectively.}
    \label{fig:transition_matrices}
\Description[Gaze transition networks for win and loss groups]{
Two directed network graphs depicting gaze transitions between functional areas of interest (AOIs) on the game interface. Each node represents a functional AOI, and directed edges represent gaze transitions between AOIs. Edge thickness encodes the magnitude of the transition probability, and node colours indicate AOI categories. The left panel (A) shows the transition network for the win group, and the right panel (B) shows the corresponding network for the loss group.
}

\end{figure}
Figure~\ref{fig:transition_matrices} shows the transition networks for each group. In both groups, gaze transitions were organised around a central hub at the \textit{Hand Cards} AOI, with the strongest transitions consistently observed from \textit{Energy} to \textit{Hand Cards}.
In the win group, \textit{Hand Cards} emerged as a stronger transition target, receiving more frequent gaze shifts from enemy-related AOIs, including both \textit{Enemy Intent} (51\%) and \textit{Enemy Status} (45\%).
In contrast, the loss group showed comparatively weaker transitions towards \textit{Hand Cards}. Instead, gaze transitions were more strongly retained within enemy-related AOIs, with a more pronounced bidirectional coupling between \textit{Enemy Intent} and \textit{Enemy Status}.
Differences were also observed in transitions involving auxiliary resources. In the win group, gaze shifts originating from \textit{Potions} were distributed across both \textit{Enemy Intent} (48\%) and \textit{Hand Cards} (26\%), whereas in the loss group these transitions were more strongly concentrated towards \textit{Enemy Intent} (56\%) and remained below the 20\% display threshold to \textit{Hand Cards}. 

\begin{figure}
    \centering
    \includegraphics[width=0.8\linewidth]{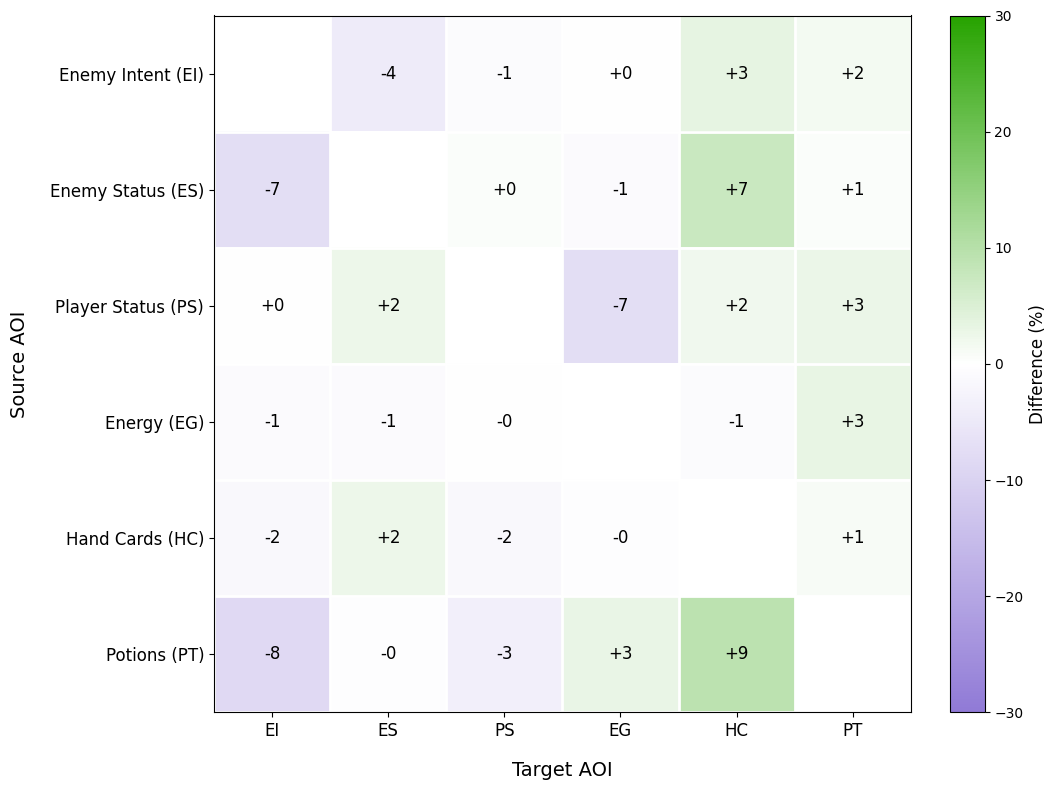}
    \caption{Difference heatmap of normalised AOI transition probabilities between win and loss groups. 
    Each cell represents the difference in transition probability from the source AOI (row) to the target AOI (column), computed as High -- Low. 
    Positive values (green) indicate transitions more frequent in the win group, while negative values (purple) indicate transitions more frequent in the loss group. 
    White cells correspond to self-transitions. 
    }
    \label{fig:aoi_diff_heatmap}
    \Description[Heatmap of differences in AOI transition probabilities between groups]{
A heatmap visualising differences in normalised AOI-to-AOI transition probabilities between win and loss groups. Rows correspond to source AOIs and columns correspond to target AOIs. Each cell shows the difference in transition probability computed as High minus Low. Green-coloured cells indicate transitions that are more frequent in the win group, while purple-coloured cells indicate transitions that are more frequent in the loss group. White cells along the diagonal represent self-transitions from an AOI to itself.
}

\end{figure}
Figure~\ref{fig:aoi_diff_heatmap} quantifies the between-group differences in transition probabilities (win minus loss). Positive values indicate transitions that were more frequent in the win group, whereas negative values indicate transitions that occurred more often in the loss group.
Notable differences were primarily observed for transitions originating from \textit{Enemy Status}, \textit{Player Status}, and \textit{Potions}. Specifically, win players showed a higher likelihood of shifting gaze from \textit{Enemy Status} to \textit{Hand Cards} ($\Delta = +7\%$), whereas loss players more frequently transitioned from \textit{Enemy Status} back to \textit{Enemy Intent} ($\Delta = -7\%$). A difference of the same magnitude was observed for transitions from \textit{Player Status}, with loss players showing a higher transition probability to \textit{Energy} ($\Delta = -7\%$). Similarly, transitions from \textit{Potions} diverged between groups, with win players more likely to redirect gaze towards \textit{Hand Cards} ($\Delta = +9\%$), while loss players showed stronger transitions from \textit{Potions} to \textit{Enemy Intent} ($\Delta = -8\%$). 

\subsubsection{Transition Matrix Entropy and Stationary Entropy}
To complement the pairwise transition analysis, we examined group differences in the broader structure of gaze transitions using two entropy measures: transition matrix entropy ($H_t^{\text{norm}}$) and stationary entropy ($H_s^{\text{norm}}$). 

\begin{figure}[!h]
    \centering
    \begin{subfigure}[t]{0.49\linewidth}
        \centering
        \includegraphics[width=0.95\linewidth]{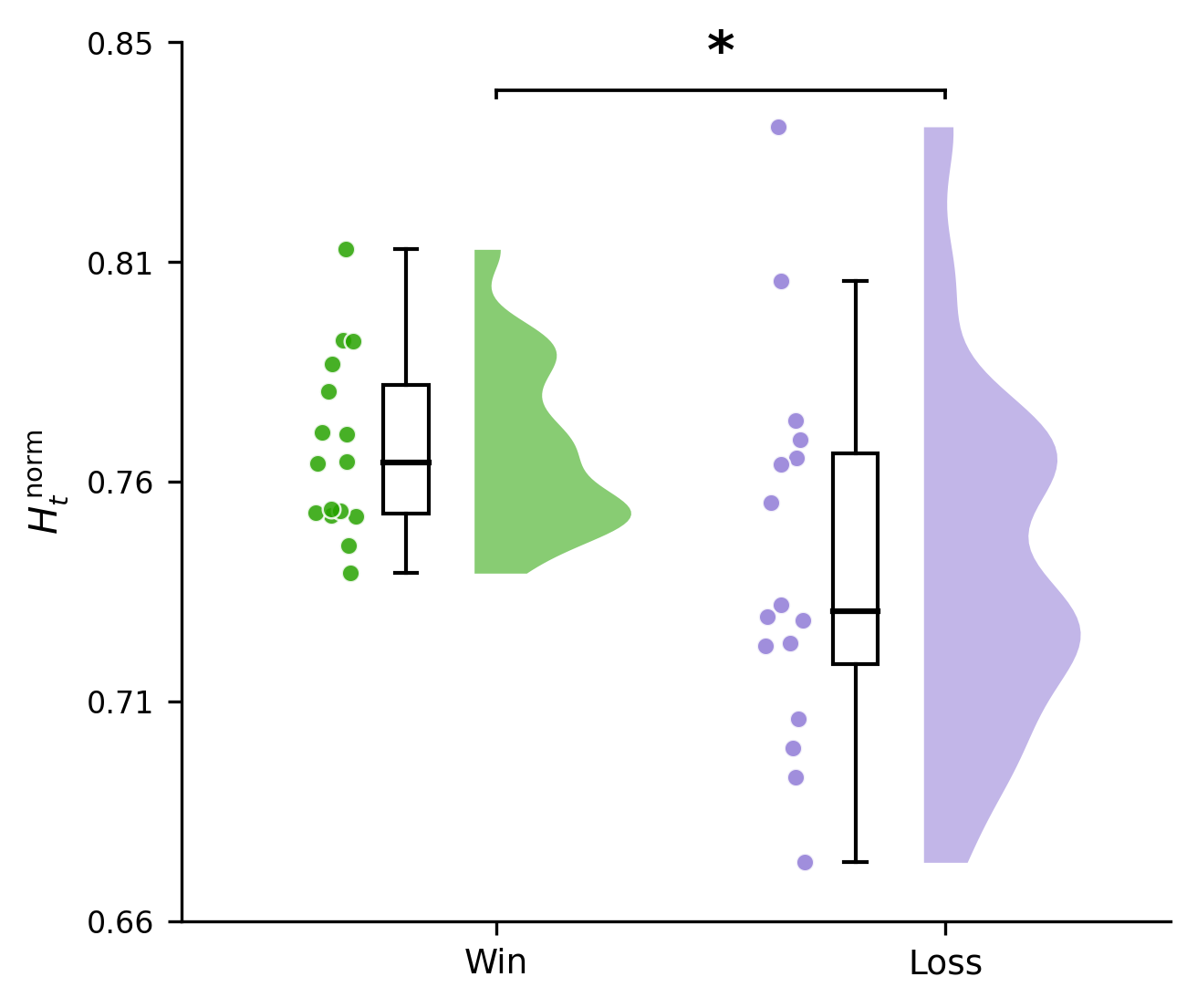}
        \caption{Transition matrix entropy ($H_t^{\,\mathrm{norm}}$).}
        \label{fig:transition_entropy_ht}
    \end{subfigure}
    \hfill
    \begin{subfigure}[t]{0.49\linewidth}
        \centering
        \includegraphics[width=0.95\linewidth]{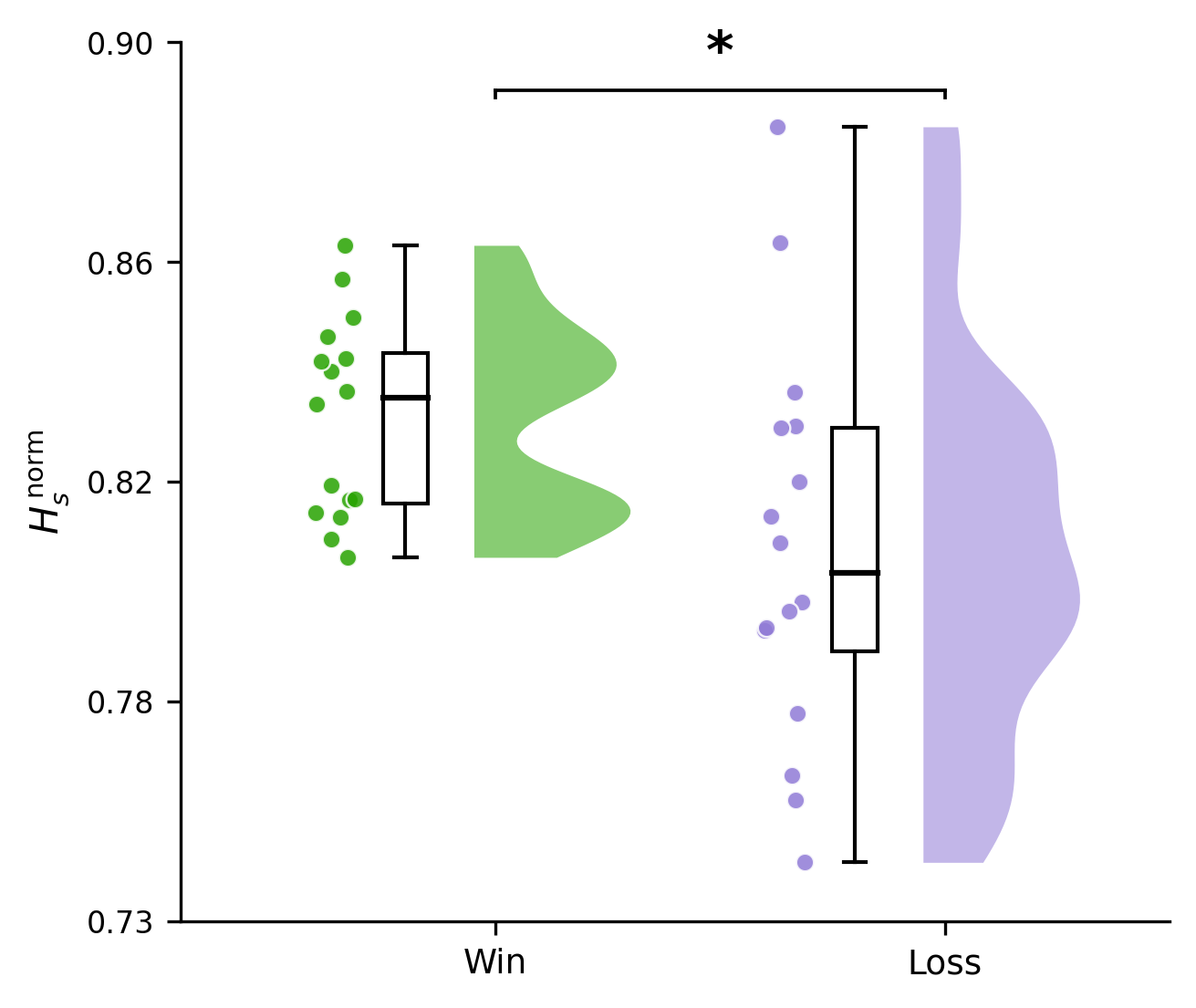}
        \caption{Stationary entropy ($H_s^{\,\mathrm{norm}}$).}
        \label{fig:transition_entropy_hs}
    \end{subfigure}
    \caption{Participant-level distributions of (a) normalised transition matrix entropy ($H_t^{\,\mathrm{norm}}$) and (b) normalised stationary entropy ($H_s^{\,\mathrm{norm}}$) across performance groups. Each point represents one participant. The half-violin shows the distribution shape, and the boxplot summarises the median and interquartile range. Asterisks indicate significant group differences ($^{*}\,p < .05$).}
    \label{fig:transition_entropy}
    \Description[Raincloud plots of normalised transition matrix entropy and stationary entropy between performance groups]{
Two raincloud plots visualising the participant-level distributions of normalised transition matrix entropy and normalised stationary entropy across performance groups. In each panel, the x-axis shows the Win and Loss groups, and the y-axis shows the corresponding normalised entropy value. Each point represents one participant. The half-violin shows the distribution shape for each group, while the boxplot summarises the median and interquartile range. In both panels, the Win group shows higher entropy values than the Loss group, with significance brackets and asterisks above the two groups indicating significant differences from Mann-Whitney U tests.
}
\end{figure}
\noindent Group differences were evaluated using two-sided Mann--Whitney U tests, and robustness was checked with permutation tests on the mean difference. The participant-level distributions are shown in Figure~\ref{fig:transition_entropy}.

For transition matrix entropy (Figure~\ref{fig:transition_entropy_ht}), the win group ($M = 0.76$, $SD = 0.02$) showed higher transition matrix entropy than the loss group ($M = 0.74$, $SD = 0.04$), yielding $\Delta H_t^{\text{norm}} = 0.02$. This difference was significant ($U = 182$, $p = .044$) with a moderate effect size ($r = 0.36$), and was corroborated by the permutation test ($\Delta H_t^{\text{norm}} = 0.02$, $p = .045$).

For stationary entropy (Figure~\ref{fig:transition_entropy_hs}), the win group ($M = 0.83$, $SD = 0.02$) exhibited higher stationary entropy than the loss group ($M = 0.80$, $SD = 0.04$), yielding $\Delta H_s^{\text{norm}} = 0.02$. This difference was significant ($U = 191$, $p = .019$) with a moderate-to-large effect size ($r = 0.42$), and was further supported by the permutation test ($\Delta H_s^{\text{norm}} = 0.02$, $p = .026$). 

\FloatBarrier
\section{Discussion}
To understand how visual attention relates to gameplay outcomes, we used eye tracking to characterise how players allocated, sustained, and shifted attention across functionally distinct game interface elements. Functional AOIs were defined based on game mechanics and gaze density maps, and were grouped into enemy information, player information, action resources, and auxiliary resources. These regions provided the basis for AOI based analyses of player attention, including hit rate, dwell time, transition probabilities, and two entropy measures. Accordingly, this discussion is organised around two aspects of gaze behaviour corresponding to our research questions. First, we examine how the distribution and duration of visual attention within functional AOIs differ between the Win and Loss groups. Second, we examine how transition patterns of visual attention between these elements differ between the two groups. Building on these findings, we then discuss design implications for supporting information to action coupling in gameplay.

\subsection{RQ1: Outcome Differences in Visual Attention Selectivity and Prioritisation}
To answer RQ1, the combined AOI hit and dwell time results indicate differences in visual attention allocation between performance groups. The Loss group exhibited longer dwell times across most AOIs, particularly enemy related information, whereas the Win group showed more selective allocation, with greater AOI hits and dwell times for auxiliary resources located in peripheral interface regions.

Among all AOIs, gaze was predominantly directed towards \textit{Hand Cards} and \textit{Enemy Intent} across both groups. Functionally, \textit{Hand Cards} are the primary site of action selection. Visually, the card area occupies a large area of the interface (see Figure~\ref{fig:density} R2), remains continuously available during combat, and updates dynamically during each turn as cards are drawn, selected, and played. These properties align with saliency-based accounts of visual attention, in which low-level features such as colour, size, motion, and intensity contribute to the selection of visually conspicuous regions \citep{itti1998model}, building on early evidence that such features are processed automatically and in parallel during preattentive vision \citep{treisman1980feature}. \textit{Hand Cards} attracted attention from both groups, reflecting both their task relevance for action selection and their stimulus-driven saliency.
\textit{Enemy Intent} represents a different source of attentional priority. Although it is not visually dominant in size, it changes across turns. Saliency-based accounts identify abrupt change and motion as low-level features that attract attention through bottom-up mechanisms \citep{itti1998model}. In addition, it conveys imminent opponent actions, making it closely tied to threat monitoring during combat. This aligns with evidence that threat-associated stimuli can guide attention even when they are not physically salient \citep{schmidt2015attentional}. 

In addition to these commonly prioritised AOIs, outcome group differences became more apparent in how players allocated attention to peripheral auxiliary resources and opponent cues.
\textit{Potions} emerged as the only AOI for which the Win group showed consistently greater attentional allocation. In the present game context, \textit{Potions} function as auxiliary resources located in a peripheral region of the combat interface. They remain visible at the edge of the visual field and provide optional, situation-dependent support rather than frequently manipulated elements of action selection during combat (see R5 in Figure~\ref{fig:aoi_diff_heatmap}).
Greater attention to this AOI in the Win group suggests proactive monitoring of peripheral resources during combat.
This observed pattern indicates an ability to take in information from peripheral regions of the interface while attending to the central action area, reflecting a wider effective span of visual attention \citep{ball1993useful}.
A similar pattern has been reported in simulated racing, where higher skilled players, grouped by faster lap times and greater weekly racing experience, allocated more attention to HUD elements outside the primary track area, allowing them to monitor task supporting information while maintaining faster performance \citep{joyce2024less}. 
This interpretation aligns with evidence from RTS gameplay showing that expert players exhibit gaze distributions spanning a larger portion of the display, enabling more rapid acquisition of visual information across the interface under high information demands \citep{jeong2022difference}.
Complementary evidence from professional eSports contexts further suggests that expert--non-expert differences are mainly reflected in the selective allocation of attention to functionally critical information sources, such as tactical overview regions in \textit{FIFA} \citep{bickmann2020gaze}.
Whereas these studies mainly compare established skill or expertise groups, the present study extends this line of work by showing that performance related selectivity can also emerge during initial exposure, before stable expertise based viewing routines have developed.

On the other hand, the Loss group exhibited longer dwell times on \textit{Enemy Intent}, a key source of information for anticipating enemy actions and guiding reactive decisions during combat.
Such prolonged fixations are typically interpreted as reflecting more effortful cognitive processing of the fixated information across reading and other information-processing tasks \citep{just1976eye, Rayner1998}. 
Similar performance-related differences in fixation duration have been reported in other game contexts. \citet{jeong2022difference} also found that lower-skill RTS players showed longer fixation durations during information acquisition, a pattern interpreted as reduced efficiency in extracting visual information. In an educational video game, low-performing players showed longer and more dispersed gaze sampling on the task-critical map region, reflecting less effective use of available cues \citep{lu2021eyes}. In the present context, the Loss group may have required more time to extract actionable meaning from enemy intent before committing to a response.

\subsection{RQ2: Outcome Differences in Integrative and Recurrent Gaze Transition Patterns}
To answer RQ2, we found that performance differences are characterised by distinct gaze transition patterns beyond isolated AOI preferences. The Win group showed stronger transitions linking auxiliary resources and enemy information to action resources, particularly Potions to Hand Cards ($\Delta = +9\%$) and Enemy Status to Hand Cards ($\Delta = +7\%$).
In contrast, the Loss group showed a stronger recurrent loop within enemy information, particularly Enemy Status to Enemy Intent ($\Delta = -7\%$) and Enemy Intent to Enemy Status ($\Delta = -4\%$).
At the level of global structure, win players exhibited higher transition and stationary entropy, indicating a broader and more even distribution of gaze across region pairs. 

Gaze transitions in the Win group more frequently linked enemy and auxiliary information to \textit{Hand Cards}, which served as the primary target for action selection. This transition pattern may indicate that the Win group more consistently translated evaluated task cues related to threat and resource monitoring into action planning, reflecting a tighter coupling between information evaluation and decision execution. Such state–action coupling aligns with the Theory of Event Coding, which describes a close linkage between perceptual information and action selection in supporting perception–action coordination \citep{hommel2001theory}.
The Loss group exhibited frequent bidirectional transitions between enemy-related AOIs (\textit{Enemy Intent} $\leftrightarrow$ \textit{Enemy Status}), forming a stronger recurrent gaze loop. This reciprocal transition pattern reflects sustained and locally recurrent sampling of enemy information and more frequent sequential switching within the same information category. According to scanpath theory \citep{noton1971scanpath}, sequences of fixations reflect how observers sample and organise visual information during interpretation. Research on text and diagram comprehension has shown that rereading gaze shifts between related information sources support the construction of an integrated mental representation \citep{hegarty1993constructing}. From this perspective, the transition pattern in our task may reflect the Loss group repeatedly sampling enemy AOIs to build an understanding of the opponent's state. Yet recurrent return to the same regions has also been linked to representations that remain incomplete: \citep{vandenbogert2014first, vandenbogert2020differences} reported that novice teachers revisited previously fixated regions when encountering teaching difficulties, a pattern described as verification of interpretation that compensates for an incomplete internal representation. The repeated return to enemy AOIs may suggest that the Loss group relied more on re-checking external cues than on working from a stable internal model of the opponent.

Beyond pairwise transitions, the two groups differed at the level of global transition structure. The Win group showed both higher transition matrix entropy and higher stationary entropy than the Loss group. In AOI-based transition analysis, higher transition entropy reflects greater variability and dispersion in gaze transitions across AOIs, whereas lower transition entropy indicates that transitions are more predictable and concentrated among a smaller subset of AOI pathways \citep{shannon1948mathematical, ellis1986statistical, cui2024gaze, shic2008statistical}. Higher stationary entropy denotes a more even distribution of gaze allocation across AOIs, while lower stationary entropy reflects stronger concentration on a limited set of AOIs \citep{krejtz2015gaze}.
Related evidence from performance based decision contexts supports the interpretation that higher gaze entropy can accompany more effective information sampling. In football refereeing research,
correct decisions were associated with higher gaze entropy, whereas more structured gaze patterns during incorrect decisions were interpreted as a possible failure to pick up situation information \citep{vanbiemen2023eyes}. Evidence from an air traffic control tracking task further showed that higher stationary gaze entropy, indicating a broader spatial distribution of gaze across the display, was associated with better response accuracy, and that high-performing novices showed higher values of both gaze transition and stationary entropy \citep{laninimaggi2021assessing}. A similar pattern has been reported in simulated aviation, where both stationary and transition entropy increased with task performance, and novice pilots in particular showed elevated entropy under more demanding conditions \citep{ayala2022characterizing}.
However, the direction of the entropy-performance relationship remains task dependent. For example, \citet{cui2024gaze} showed that both stationary entropy and transition entropy decreased when participants had to maintain attention on a demanding tracking component, but increased when communication prompts occurred more frequently and required gaze to be redistributed across multiple task areas. Entropy therefore needs to be interpreted in relation to the task structure.
In the present first-time gameplay setting, the higher entropy observed in the Win group aligned with their greater hit rate and dwell time on peripheral auxiliary resources, and with the broader set of transition pathways in the matrix analysis. This convergence suggests that the elevated entropy in the Win group reflected task-driven coverage of decision-relevant information rather than random or indiscriminate scanning. 
The relevance of broader gaze structure can be understood through situation awareness, where effective decisions in dynamic systems depend on perceiving relevant elements, understanding their meaning, and using that understanding to guide subsequent actions \citep{endsley1995toward}. In an early exposure context, 
broader sampling across information sources may have helped players form situational understanding without extensive prior experience. In contrast, lower entropy in the Loss group reflects a more constrained transition landscape, which may have reduced access to contextual information needed to support effective decisions. Furthermore, the role of entropy may change as game familiarity develops, and future longitudinal work is needed to examine how this relationship evolves over repeated play.

\subsection{Design Implications: Supporting Information-to-Action Coupling}


Within a turn-based game setting, our results point to a design opportunity concerning how players translate task-relevant information into action. One pattern concerns peripheral auxiliary resources. Loss players allocated fewer hits and shorter dwell times to Potions, and their gaze transitions from this region toward action resources fell below the display threshold in the transition network, while transitions from Potions to Enemy Intent were comparatively strong ($\Delta = -8\%$). These patterns together indicate that peripheral resources were sampled but not consistently integrated into action selection: because a potion's benefit can only be realised when it is evaluated together with the cards available to play, the tendency to return gaze to threat monitoring rather than to action resources suggests that an available resource was registered but not brought into the turn's action planning. A complementary pattern emerged from enemy information. Loss players exhibited a stronger recurrent loop between Enemy Intent and Enemy Status, with comparatively weak transitions from these regions toward Hand Cards, whereas win players more often redirected gaze from enemy information to action resources (Enemy Status → Hand Cards, $\Delta = +7\%$). Given that key threat information remains visible during a turn, the loop is unlikely to reflect a difficulty in accessing enemy cues alone, and may instead reflect a tendency for assessed threats not to be consistently carried forward into a response. Together, loss players showed under-sampling of a peripheral resource; on the other hand, they showed over-sampling of threat information. In both cases, gaze reached the relevant information but did not consistently flow onward to action selection. The limitation lies less in what was attended than in the coordination of information into a coherent internal representation from which an action can be selected.

Prior approaches to directing visual attention in games have sought to make important elements more conspicuous by automatically adjusting stimulus-driven manipulations such as lighting, contrast, or brightness, raising the bottom-up salience of a target \citep{SeifElNasr2009ALVA}. Our findings complement this direction in turn-based decision contexts, where salience-based support could be complemented by interventions that integrate perceived cues into the flow of decision making once they are perceived.
We therefore suggest that interfaces could help players connect perceived cues to feasible actions, particularly when the current game state makes those actions decision relevant. For example, when an enemy signals an imminent high-damage action, the cards in hand that can mitigate it could be marked with a brief outline. Similarly, when a beneficial potion is available and its effect is pertinent to the present situation, the corresponding slot could briefly indicate its actionable effect. The hand cards whose use it complements could be concurrently marked, externalising the link between the resource and a viable course of action. This design approach aligns with the Attentional Capture assumption \citep{simons2000attentional}, which emphasises that stimuli explicitly aligned with task objectives or meaningful goals are more likely to attract attention effectively. Because such support is triggered by an inferred gaze pattern and a specific game state rather than by a pre-specified importance value, it can be applied selectively to the game situations in which integration is most likely to break down. In doing so, the interface could preserve player agency while avoiding persistent visual clutter. More broadly, transition based measures could offer a way to evaluate whether an intervention genuinely restructures a player’s gaze pattern rather than merely attracting momentary attention.

\section{Limitations and Future Work}
We identified five key limitations in the interpretation of the present findings. These relate to (1) AOI extraction accuracy and consistency under dynamic combat layouts, (2) the outcome-based definition of performance groups, (3) the scope of the performance group comparison, (4) variation in hand-card availability, and (5) task specificity of gaze-based findings.

The first limitation concerns the extraction of AOI positions. Although the enemy types in the game are limited, their spatial arrangements vary considerably across encounters, particularly depending on the number of enemies presented on screen. To address this, we defined and extracted separate AOI templates for different encounters to accommodate distinct spatial layouts and phase-dependent visual differences. However, residual inconsistencies may still exist within each template. Even among encounters of the same type, particularly under multi-enemy conditions, the number of active enemies, their spacing, relative sizes, and animation frames can slightly alter the position of visual elements, resulting in alignment errors between AOI boundaries and actual UI features. Moreover, as enemies appear or are defeated during combat, the spatial correspondence between predefined AOIs and specific opponents can dynamically change. This dynamic change was reflected in our template alignment validation, with the Enemy Intent AOI showing a within-template alignment rate of 82\%. Such within-template variability may introduce noise to gaze-based measures. Future work could address these issues by implementing an adaptive mapping approach for AOI extraction. Instead of relying on predefined static templates, future studies could employ machine-learning-based tracking to automatically detect and update the positions of relevant interface elements in real time. Techniques such as feature-based image registration or object detection networks could continuously align AOI boundaries with the current interface layout as it changes during gameplay.

A second limitation concerns the outcome-based definition of performance groups. The win and loss groups were defined by the final win/loss outcome of the run, rather than by an independently measured skill level or experimentally manipulated performance condition. In a deck-building game, this outcome can reflect multiple interacting factors across the run, including card choices, deck development, resource use, and combat action sequences. Therefore, the observed group differences should be interpreted as gaze patterns associated with naturally emerging gameplay outcomes, rather than as differences attributable solely to visual attention strategies.

A third limitation concerns the scope of the performance group comparison. The experiment was designed to capture visual attention across a full gameplay progression, from an earlier stage to a later boss fight with increased demands. However, the comparison was still based on participants’ initial exposure to the game within one experimental playthrough. As visual attention patterns may evolve with learning and growing familiarity, the observed differences should be interpreted as early outcome related differences rather than stable performance characteristics. Future work could employ a longitudinal design to examine the consistency of gaze behaviour across repeated play. This would help clarify whether the observed gaze patterns reflect stable performance related differences. In addition, future research could investigate temporal trends under increasing task demands by tracking how visual attention shifts across gameplay progression. Such analysis may offer a more detailed account of when differences between performance groups begin to emerge and how attentional strategies adapt as difficulty intensifies. 

A fourth limitation concerns turn level variation in hand card availability. The current study used a fixed character and constant run seed to standardise the map layout and enemy sequence across participants. However, the specific cards available to each player could still vary with deck composition, prior choices, and action sequences during gameplay. Attention to \textit{Hand Cards} may therefore be partly shaped by the cost, usability, and tactical relevance of the cards currently in hand. Future work could combine gaze data with card level information. Instead of treating \textit{Hand Cards} as a single AOI, future studies could integrate card variables, such as card type, card cost, hand composition, and card use history, to analyse how players inspect individual cards, shift gaze across card options, and evaluate and sequence card plays during gameplay. Such an analysis would move beyond the present region level AOI analysis and enable a more detailed examination of card based action selection.

A fifth limitation concerns the task specificity of the study context. Our study was conducted within a single game environment, characterised by a relatively consistent combat interface and structured decision-making mechanics. However, gaze-based measures are highly sensitive to task structure and visual layout, and the functional definition of AOIs and their associated information demands is inherently task-specific. Accordingly, the visual attention patterns observed in this study should be interpreted as context-dependent, and their generalisability to other game genres or interactive systems with different UI complexity remains to be validated. Future work should examine whether performance-related differences in visual attention generalise to broader interactive contexts, such as virtual learning and performance settings \citep{rappa2022use}, in order to establish the boundary conditions under which distinct gaze allocation and transition patterns remain discriminative of performance.



\section{Conclusions}
This study investigated the distribution, duration, and transition of visual attention in relation to gameplay outcomes in a turn-based deck-building game. 
We defined AOIs through a hybrid procedure that combined functional categorisation of the game interface with gaze density refinement, and computed AOI based measures of gaze allocation, dwell time, transitions, and entropy.
Players in the win group showed a more selective distribution of visual attention, with greater attention to peripheral auxiliary resources, whereas players in the loss group showed longer dwell time on enemy information.
Gaze transition analysis further showed that players in the win group used a broader set of directed AOI transitions, with higher transition complexity and a more even distribution of attention across AOIs.
These findings indicate that gameplay outcomes are reflected in where attention is directed, how long it is sustained, and how it shifts between functional information sources during play.
From a design perspective, our results suggest that interfaces could help players link attended information to the actions available in the current game state, externalising the connection between situational cues and viable responses. These findings point to gaze-based approaches that consider not only where attention is directed but also how it is coordinated with decision making during gameplay.
Future work could refine gaze-based analyses through adaptive AOI extraction that accommodates dynamic interface layouts, longitudinal designs that track how visual attention develops with familiarity and increasing task demands, and finer grained analyses that integrate gaze patterns with action level information, such as card selection. Further validation across different game contexts would also help establish the conditions under which these gaze patterns generalise.

\bibliographystyle{ACM-Reference-Format}
\bibliography{main}

@incollection{zammitto2014gaming,
  title     = {Gaming},
  author    = {Zammitto, Veronica and Steiner, Karl},
  booktitle = {Eye Tracking in User Experience Design},
  editor    = {Bergstrom, Jennifer Romano and Schall, Andrew Jonathan},
  chapter   = {11},
  pages     = {291--310},
  year      = {2014},
  publisher = {Morgan Kaufmann},
  doi       = {10.1016/B978-0-12-408138-3.00011-X}
}

@article{vandenbogert2014first,
  author  = {van den Bogert, Neeske and van Bruggen, Jan and Kostons, Danny and Jochems, Wim},
  title   = {First Steps Into Understanding Teachers’ Visual Perception of Classroom Events},
  journal = {Teaching and Teacher Education},
  volume  = {37},
  pages   = {208--216},
  year    = {2014},
  doi     = {10.1016/j.tate.2013.09.001}
}

@article{hommel2001theory,
  author  = {Hommel, Bernhard and M{\"u}sseler, Jochen and Aschersleben, Gisa and Prinz, Wolfgang},
  title   = {The Theory of Event Coding (TEC): A Framework for Perception and Action Planning},
  journal = {Behavioral and Brain Sciences},
  volume  = {24},
  number  = {5},
  pages   = {849--937},
  year    = {2001},
  doi     = {10.1017/S0140525X01000103}
}

@article{simons2000attentional,
  title={Attentional capture and inattentional blindness},
  author={Simons, Daniel J},
  journal={Trends in cognitive sciences},
  volume={4},
  number={4},
  pages={147--155},
  year={2000},
  publisher={Elsevier}
}

@incollection{wickens1995multiple,
  title={Multiple resources and performance prediction},
  author={Wickens, Christopher D.},
  booktitle={Handbook of Perception and Human Performance},
  editor={Boff, Kenneth R. and Kaufman, Lloyd and Thomas, James P.},
  volume={2},
  pages={71--1},
  year={1995},
  publisher={Wiley},
  address={New York}
}

@misc{SlayWiki,
  author       = {{MegaCrit}},
  title        = {Slay the Spire Wiki},
  howpublished = {\url{https://slay-the-spire.fandom.com/wiki/Slay_the_Spire_Wiki}},
  note         = {Accessed: October 6, 2025},
  year         = {2025},
  organization = {Fandom},
}

@article{krejtz2015gaze,
  title        = {Gaze Transition Entropy},
  author       = {Krejtz, Krzysztof and Duchowski, Andrew T. and Szmidt, Tomasz and Krejtz, Izabela and González Perilli, Fernando and Pires, Ana and Vilar, Eva and Villalobos, Nicolas},
  journal      = {ACM Transactions on Applied Perception},
  volume       = {13},
  number       = {1},
  pages        = {1--20},
  year         = {2015},
  publisher    = {Association for Computing Machinery},
  doi          = {10.1145/2842602}
}

@article{shiferaw2019review,
  title={A review of gaze entropy as a measure of visual scanning efficiency},
  author={Shiferaw, Brook and Downey, Luke and Crewther, David},
  journal={Neuroscience \& Biobehavioral Reviews},
  volume={96},
  pages={353--366},
  year={2019},
  publisher={Elsevier}
}

@article{noton1971scanpath,
  title        = {Scanpaths in Eye Movements during Pattern Perception},
  author       = {Noton, David and Stark, Lawrence},
  journal      = {Science},
  volume       = {171},
  number       = {3968},
  pages        = {308--311},
  year         = {1971},
  publisher    = {American Association for the Advancement of Science},
  doi          = {10.1126/science.171.3968.308}
}

@article{vandenbogert2020differences,
  author  = {van den Bogert, Neeske and van Bruggen, Jan and Kostons, Danny and Jochems, Wim},
  title   = {Differences in Eye Movements Between Experts and Novices When Viewing Teacher--Student Interactions},
  journal = {Frontiers in Education},
  volume  = {5},
  pages   = {612175},
  year    = {2020},
  doi     = {10.3389/feduc.2020.612175}
}

@article{conati2013understanding,
  title={Understanding attention to adaptive hints in educational games: An eye-tracking study},
  author={Conati, Cristina and Jaques, Natasha and Muir, Michelle},
  journal={International Journal of Artificial Intelligence in Education},
  volume={23},
  number={1},
  pages={136--161},
  year={2013},
  doi={10.1007/s40593-013-0002-8}
}

@book{kahneman1973attention,
  author    = {Kahneman, Daniel},
  title     = {Attention and Effort},
  year      = {1973},
  publisher = {Prentice-Hall},
  address   = {Englewood Cliffs, NJ}
}

@article{treue2003visual,
  author  = {Treue, Stefan},
  title   = {Visual Attention: The Where, What, How and Why of Saliency},
  journal = {Current Opinion in Neurobiology},
  year    = {2003},
  volume  = {13},
  number  = {4},
  pages   = {428--432},
  doi     = {10.1016/S0959-4388(03)00105-3}
}

@article{carrasco2011visual,
  author  = {Carrasco, Marisa},
  title   = {Visual Attention: The Past 25 Years},
  journal = {Vision Research},
  year    = {2011},
  volume  = {51},
  number  = {13},
  pages   = {1484--1525},
  doi     = {10.1016/j.visres.2011.04.012}
}

@inproceedings{kirillov2023segment,
title={Segment Anything},
author={Kirillov, Alexander and Mintun, Eric and Ravi, Nikhila and Mao, Hanzi and Rolland, Chloe and Gustafson, Laura and Xiao, Tete and Whitehead, Spencer and Berg, Alexander C. and Lo, Wan-Yen and Doll{'a}r, Piotr and Girshick, Ross},
booktitle={Proceedings of the IEEE/CVF International Conference on Computer Vision},
pages={4015--4026},
year={2023}
}

@article{ravi2024sam2,
title={SAM 2: Segment Anything in Images and Videos},
author={Ravi, Nikhila and Gabeur, Valentin and Hu, Yuan-Ting and Hu, Ronghang and Ryali, Chaitanya and Ma, Tengyu and Khedr, Haitham and R{"a}dle, Roman and Rolland, Chloe and Gustafson, Laura and Mintun, Eric and Pan, Junting and Alwala, Kalyan Vasudev and Carion, Nicolas and Wu, Chao-Yuan and Girshick, Ross and Doll{'a}r, Piotr and Feichtenhofer, Christoph},
journal={arXiv preprint arXiv:2408.00714},
year={2024}
}

@article{treisman1980feature,
  title={A feature-integration theory of attention},
  author={Treisman, Anne M. and Gelade, Garry},
  journal={Cognitive Psychology},
  volume={12},
  number={1},
  pages={97--136},
  year={1980},
  publisher={Elsevier},
  doi={10.1016/0010-0285(80)90005-5}
}

@article{schmidt2015attentional,
  title={Attentional capture by signals of threat},
  author={Schmidt, Lisette J and Belopolsky, Artem V and Theeuwes, Jan},
  journal={Cognition and emotion},
  volume={29},
  number={4},
  pages={687--694},
  year={2015},
  publisher={Taylor \& Francis}
}

@article{itti1998model,
  title={A model of saliency-based visual attention for rapid scene analysis},
  author={Itti, Laurent and Koch, Christof and Niebur, Ernst},
  journal={IEEE Transactions on pattern analysis and machine intelligence},
  volume={20},
  number={11},
  pages={1254--1259},
  year={1998},
  publisher={Ieee}
}

@inproceedings{Chung2017TheIO,
	title        = {The Impact of Visual Load on Performance in a Human-Computation Game},
	author       = {Christina Chung and Amit Kadan and Yueti Yang and Asako Matsuoka and Julia Rubin and Marsha Chechik},
	year         = 2017,
	booktitle    = {Proceedings of the 12th International Conference on the Foundations of Digital Games},
	doi          = {10.1145/3102071.3106358}
}

@article{bickmann2020gaze,
  title={Gaze behavior of professional and non-professional esports players in FIFA 19},
  author={Bickmann, Peter and Wechsler, Konstantin and Rudolf, Kevin and Tholl, Chuck and Frob{\"o}se, Ingo and Grieben, Christopher},
  journal={International Journal of Gaming and Computer-Mediated Simulations (IJGCMS)},
  volume={12},
  number={3},
  pages={1--17},
  year={2020},
  publisher={IGI Global Scientific Publishing}
}

@article{jeong2022difference,
  title={Difference in gaze control ability between low and high skill players of a real-time strategy game in esports},
  author={Jeong, Inhyeok and Nakagawa, Kento and Osu, Rieko and Kanosue, Kazuyuki},
  journal={PloS one},
  volume={17},
  number={3},
  pages={e0265526},
  year={2022},
  publisher={Public Library of Science San Francisco, CA USA}
}

@article{gotardi2019combining,
  title={Combining experiences of race gaming and natural driving affects gaze location strategy in simulated context},
  author={Gotardi, Gisele C and Kuga, Gabriel K and Sim{\~a}o, Rafael O and Brito, Matheus B and Paschoalino, Gabriel P and Silva, Gustavo A and Barbieri, Fabio A and Polastri, Paula F and Schor, Paulo and Navarro, Martina},
  journal={Ergonomics},
  volume={62},
  number={11},
  pages={1392--1399},
  year={2019},
  publisher={Taylor \& Francis}
}

@article{joyce2024less,
  title = {Less Is More: Higher-Skilled Sim Racers Allocate Significantly Less Attention to the Track Relative to the Display Features than Lower-Skilled Sim Racers},
  author = {Joyce, John M. and Campbell, Mark J. and T{\'o}th, {\'A}d{\'a}m J. and Hojaji, Fazilat},
  journal = {Vision},
  volume = {8},
  number = {2},
  pages = {27},
  year = {2024},
  doi = {10.3390/vision8020027}
}

@inproceedings{shic2008statistical,
  title     = {A statistical approach to the analysis of eye-tracking data},
  author    = {Shic, Frederick and Chawarska, Katarzyna and Scassellati, Brian},
  booktitle = {Proceedings of the 7th IEEE International Conference on Development and Learning (ICDL)},
  year      = {2008},
  pages     = {1--6},
  publisher = {IEEE},
  doi       = {10.1109/DEVLRN.2008.4640840}
}

@inproceedings{el2006visual,
  title={Visual attention in 3D video games},
  author={El-Nasr, Magy Seif and Yan, Su},
  booktitle={Proceedings of the 2006 ACM SIGCHI international conference on Advances in computer entertainment technology},
  pages={22--es},
  year={2006}
}

@article{SeifElNasr2009ALVA,
  author  = {Seif El-Nasr, Magy and Vasilakos, Athanasios and Rao, Chinmay and Zupko, Joseph},
  title   = {Dynamic Intelligent Lighting for Directing Visual Attention in Interactive 3{D} Scenes},
  journal = {IEEE Transactions on Computational Intelligence and AI in Games},
  volume  = {1},
  number  = {2},
  pages   = {145--153},
  year    = {2009},
}

@article{almeida2016video,
  title={Video game scenery analysis with eye tracking},
  author={Almeida, Samuel and Mealha, {\'O}scar and Veloso, Ana},
  journal={Entertainment Computing},
  volume={14},
  pages={1--13},
  year={2016},
  publisher={Elsevier}
}

@article{rappa2022use,
  title={The use of eye tracking technology to explore learning and performance within virtual reality and mixed reality settings: A scoping review},
  author={Rappa, Natasha Anne and Ledger, Susan and Teo, Timothy and Wong, Kok Wai and Power, Brad and Hilliard, Bruce},
  journal={Interactive Learning Environments},
  volume={30},
  number={7},
  pages={1338--1350},
  year={2022},
  publisher={Taylor \& Francis},
  doi={10.1080/10494820.2019.1702560}
}

@inproceedings{almeida2011eyes,
  title={The Eyes and Games: A Survey of Visual Attention and Eye Tracking Input in Video Games},
  author={Almeida, Samuel and Veloso, Ana and Roque, Lic{\'i}nio and Mealha, {\'O}scar},
  booktitle={Proceedings of SBGames 2011: X Brazilian Symposium on Computer Games and Digital Entertainment},
  pages={1--10},
  year={2011}
}

@book{cheng2014relationship,
  title={Relationship between visual attention and flow experience in a serious educational game: An eye tracking analysis},
  author={Cheng, Wai Ki Rebecca},
  year={2014},
  publisher={George Mason University}
}

@book{sundstedt2012gazing,
  title={Gazing at games: An introduction to eye tracking control},
  author={Sundstedt, Veronica},
  volume={14},
  year={2012},
  publisher={Morgan \& Claypool Publishers}
}

@inproceedings{mat2011eye,
  title={Eye tracking in educational games environment: evaluating user interface design through eye tracking patterns},
  author={Mat Zain, Nurul Hidayah and Abdul Razak, Fariza Hanis and Jaafar, Azizah and Zulkipli, Mohd Firdaus},
  booktitle={International Visual Informatics Conference},
  pages={64--73},
  year={2011},
  organization={Springer}
}

@inproceedings{Sundstedt2008FixationBehavior,
	title        = {A Psychophysical Study of Fixation Behavior in a Computer Game},
	author       = {Veronica Sundstedt and Efstathios Stavrakis and Michael Wimmer and Erik Reinhard},
	year         = 2008,
	booktitle    = {Proceedings of the 5th Symposium on Applied Perception in Graphics and Visualization},
	publisher    = {ACM},
	pages        = {43--50},
	doi          = {10.1145/1394281.1394288}
}

@article{opencv_library,
	title        = {The OpenCV Library},
	author       = {Garry Bradski},
	year         = 2000,
	journal      = {Dr. Dobb's Journal of Software Tools}
}

@article{vanbiemen2023eyes,
  author  = {van Biemen, T. and Oudejans, R. R. D. and Savelsbergh, G. J. P. and Zwenk, F. and Mann, D. L.},
  title   = {Into the Eyes of the Referee: A Comparison of Elite and Sub-Elite Football Referees' On-Field Visual Search Behaviour when Making Foul Judgements},
  journal = {International Journal of Sports Science \& Coaching},
  volume  = {18},
  number  = {1},
  pages   = {78--90},
  year    = {2023},
  doi     = {10.1177/17479541211069469}
}

@article{ball1993useful,
  author    = {Ball, Karlene and Owsley, Cynthia},
  title     = {The useful field of view test: A new technique for evaluating age-related declines in visual function},
  journal   = {Journal of the American Optometric Association},
  year      = {1993},
  volume    = {64},
  number    = {1},
  pages     = {71--79}
}

@article{ayala2022characterizing,
  author    = {Ayala, Naila and Kearns, Sonya and Cao, Suzanne and Irving, Elizabeth and Niechwiej-Szwedo, Ewa},
  title     = {Characterizing individual differences in task performance and task difficulty with gaze entropy},
  journal   = {Journal of Vision},
  year      = {2022},
  volume    = {22},
  number    = {14},
  pages     = {3692},
  doi       = {10.1167/jov.22.14.3692}
}

@article{ellis1986statistical,
  author    = {Ellis, Stephen R. and Stark, Lawrence},
  title     = {Statistical Dependency in Visual Scanning},
  journal   = {Human Factors},
  year      = {1986},
  volume    = {28},
  number    = {4},
  pages     = {421--438},
  doi       = {10.1177/001872088602800405}
}

@article{cui2024gaze,
  author    = {Cui, Yuhao and Sato, Toyomi and Crundall, David and Trick, Lana M.},
  title     = {Gaze transition entropy as a measure of attention allocation in a dynamic workspace involving automation},
  journal   = {Scientific Reports},
  year      = {2024},
  volume    = {14},
  pages     = {23488},
  doi       = {10.1038/s41598-024-74244-4}
}

@article{laninimaggi2021assessing,
  author  = {Lanini-Maggi, Sara and Ruginski, Ian T. and Shipley, Thomas F. and Hurter, Christophe and Duchowski, Andrew T. and Briesemeister, Benny B. and Lee, Jihyun and Fabrikant, Sara I.},
  title   = {Assessing How Visual Search Entropy and Engagement Predict Performance in a Multiple-Objects Tracking Air Traffic Control Task},
  journal = {Computers in Human Behavior Reports},
  volume  = {4},
  pages   = {100127},
  year    = {2021},
  doi     = {10.1016/j.chbr.2021.100127}
}

@article{endsley1995toward,
	title        = {Toward a theory of situation awareness in dynamic systems},
	author       = {Endsley, Mica R},
	year         = 1995,
	journal      = {Human factors},
	publisher    = {SAGE Publications Sage CA: Los Angeles, CA},
	volume       = 37,
	number       = 1,
	pages        = {32--64}
}

@article{jordan2009analysis,
	title        = {An analysis of eye scanpath entropy in a progressively forming virtual environment},
	author       = {Jordan, Joel and Slater, Mel},
	year         = 2009,
	journal      = {Presence},
	publisher    = {MIT Press},
	volume       = 18,
	number       = 3,
	pages        = {185--199}
}

@article{shannon1948mathematical,
	title        = {A Mathematical Theory of Communication},
	author       = {Shannon, Claude E.},
	year         = 1948,
	journal      = {Bell System Technical Journal},
	volume       = 27,
	number       = 3,
	pages        = {379--423},
	doi          = {10.1002/j.1538-7305.1948.tb01338.x}
}

@article{roda2011human,
  title={Human attention and its implications for human--computer interaction},
  author={Roda, Claudia},
  journal={Human attention in digital environments},
  volume={1},
  pages={11--62},
  year={2011},
  publisher={Cambridge University Press Cambridge}
}

@incollection{green2002where,
  author    = {Green, Paul},
  title     = {Where Do Drivers Look While Driving (and for How Long)?},
  booktitle = {Human Factors in Traffic Safety},
  editor    = {Dewar, Robert E. and Olson, Paul L.},
  pages     = {77--110},
  year      = {2002},
  publisher = {Lawyers \& Judges},
  address   = {Tucson, AZ}
}

@inbook{poole2006eye,
  title     = {Eye Tracking in HCI and Usability Research},
  author    = {Poole, Alex and Ball, Linden J.},
  editor    = {Ghaoui, Claude},
  booktitle = {Encyclopaedia of Human-Computer Interaction},
  pages     = {211--219},
  publisher = {Idea Group Inc.},
  address   = {Pennsylvania},
  year      = {2006}
}

@article{wang2024comparative,
  title={A comparative analysis for eye movement characteristics between professional and non-professional players in FIFA eSports game},
  author={Wang, Haoyue and Yang, Jian and Hu, Menghan and Tang, Jingyu and Yu, Wangyang},
  journal={Displays},
  volume={81},
  pages={102599},
  year={2024},
  publisher={Elsevier}
}

@misc{slaythespire,
	title        = {Slay the Spire},
	author       = {{MegaCrit}},
	year         = 2017,
	publisher    = {MegaCrit},
	howpublished = {Video Game, Available on PC, PlayStation 4, Xbox One, Nintendo Switch}
}

@article{luo2025differences,
  title={Differences in eye movement characteristics between expert and non-expert eSports players: a systematic review and meta-analysis},
  author={Luo, Yekui and Chen, Yonghuan and Cho, Jinho and Yan, Changliang and Seo, JaeChul},
  journal={Scientific Reports},
  volume={15},
  number={1},
  pages={30185},
  year={2025},
  publisher={Nature Publishing Group UK London}
}

@book{holmqvist2011eye,
	title        = {Eye Tracking: A Comprehensive Guide to Methods and Measures},
	author       = {Holmqvist, Kenneth and Nystr{\"o}m, Marcus and Andersson, Richard and Dewhurst, Richard and Jarodzka, Halszka and Van de Weijer, Joost},
	year         = 2011,
	publisher    = {Oxford University Press}
}

@inproceedings{lu2021eyes,
  title={What the eyes can tell: Analyzing visual attention with an educational video game},
  author={Lu, Wenyi and He, Hao and Urban, Alex and Griffin, Joe},
  booktitle={ACM Symposium on Eye Tracking Research and Applications},
  pages={1--7},
  year={2021}
}

@incollection{sundstedt2013visual,
  title={Visual attention and gaze behavior in games: An object-based approach},
  author={Sundstedt, Veronica and Bernhard, Matthias and Stavrakis, Efstathios and Reinhard, Erik and Wimmer, Michael},
  booktitle={Game analytics: Maximizing the value of player data},
  pages={543--583},
  year={2013},
  publisher={Springer}
}

@book{jorgensen2013gameworld,
  title={Gameworld interfaces},
  author={Jorgensen, Kristine},
  year={2013},
  publisher={MIT Press}
}

@inproceedings{llanos2011players,
  title={Do players prefer integrated user interfaces? A qualitative study of game UI design issues},
  author={Llanos, Stein C and J{\o}rgensen, Kristine},
  booktitle={Proceedings of DiGRA 2011 Conference: Think Design Play},
  year={2011}
}

@book{duchowski2017eye,
  title={Eye Tracking Methodology: Theory and Practice},
  author={Duchowski, Andrew T.},
  edition={3rd},
  year={2017},
  publisher={Springer},
  address={Cham, Switzerland},
  doi={10.1007/978-3-319-57883-5}
}

@article{Rayner1998,
  title={Eye movements in reading and information processing: 20 years of research},
  author={Rayner, Keith},
  journal={Psychological Bulletin},
  volume={124},
  number={3},
  pages={372--422},
  year={1998},
  publisher={American Psychological Association},
  doi={10.1037/0033-2909.124.3.372}
}

@article{negi2020fixation,
  title={Fixation duration and the learning process: An eye tracking study with subtitled videos},
  author={Negi, Shivsevak and Mitra, Ritayan},
  journal={Journal of Eye Movement Research},
  volume={13},
  number={6},
  pages={1--15},
  year={2020},
  publisher={Bern Open Publishing}
}

@article{jiang2019applying,
  title={Applying eye-tracking technology to measure interactive experience toward the navigation interface of mobile games considering different visual attention patterns},
  author={Jiang, Jiayan and Guo, Feng and Chen, Junhua and Tian, Xuehua and Lv, Wei},
  journal={Applied Sciences},
  volume={9},
  number={19},
  pages={4076},
  year={2019},
  publisher={MDPI},
  doi={10.3390/app9194076}
}

@article{yu2024effects,
  title={Effects of player--video game interaction on the mental effort of older adults with the use of electroencephalography and NASA-TLX},
  author={Yu, RWL and Chan, AHS},
  journal={Archives of Gerontology and Geriatrics},
  volume={124},
  pages={105442},
  year={2024},
  publisher={Elsevier}
}

@article{corbetta2002control,
  author  = {Corbetta, Maurizio and Shulman, Gordon L.},
  title   = {Control of Goal-Directed and Stimulus-Driven Attention in the Brain},
  journal = {Nature Reviews Neuroscience},
  year    = {2002},
  volume  = {3},
  number  = {3},
  pages   = {201--215},
  doi     = {10.1038/nrn755}
}

@article{desimone1995neural,
  author  = {Desimone, Robert and Duncan, John},
  title   = {Neural Mechanisms of Selective Visual Attention},
  journal = {Annual Review of Neuroscience},
  year    = {1995},
  volume  = {18},
  pages   = {193--222},
  doi     = {10.1146/annurev.ne.18.030195.001205}
}

@article{10.1145/3352763,
author = {Hadnett-Hunter, Jacob and Nicolaou, George and O'Neill, Eamonn and Proulx, Michael},
title = {The Effect of Task on Visual Attention in Interactive Virtual Environments},
year = {2019},
issue_date = {July 2019},
publisher = {Association for Computing Machinery},
address = {New York, NY, USA},
volume = {16},
number = {3},
issn = {1544-3558},
url = {https://doi-org.virtual.anu.edu.au/10.1145/3352763},
doi = {10.1145/3352763},
journal = {ACM Trans. Appl. Percept.},
month = sep,
articleno = {17},
numpages = {17}
}

@article{wolfe1994guided,
  author  = {Wolfe, Jeremy M.},
  title   = {Guided Search 2.0: A Revised Model of Visual Search},
  journal = {Psychonomic Bulletin \& Review},
  year    = {1994},
  volume  = {1},
  number  = {2},
  pages   = {202--238},
  doi     = {10.3758/BF03200774}
}

@book{yarbus1967eye,
  author    = {Yarbus, Alfred L.},
  title     = {Eye Movements and Vision},
  year      = {1967},
  publisher = {Plenum Press},
  address   = {New York}
}

@inproceedings{oyekoya2009saliency,
  title={A saliency-based method of simulating visual attention in virtual scenes},
  author={Oyekoya, Oyewole and Steptoe, William and Steed, Anthony},
  booktitle={Proceedings of the 16th ACM symposium on virtual reality software and technology},
  pages={199--206},
  year={2009}
}

@misc{TobiiProFusion,
  author       = {{Tobii AB}},
  title        = {Tobii Pro Fusion Eye Tracker},
  howpublished = {\url{https://www.tobii.com/products/eye-trackers/screen-based/tobii-pro-fusion}},
  note         = {Accessed: 2025-09-10},
  year         = {2019}
}

@misc{tobii_accuracy_precision,
  author       = {{Tobii Technology AB}},
  title        = {Let's talk accuracy and precision},
  year         = {2023},
  url          = {https://help.tobii.com/hc/en-us/articles/213534825-Let-s-talk-accuracy-and-precision},
  note         = {Accessed: 2025-09-07}
}

@article{hegarty1993constructing,
  title={Constructing mental models of machines from text and diagrams},
  author={Hegarty, Mary and Just, Marcel Adam},
  journal={Journal of memory and language},
  volume={32},
  number={6},
  pages={717--742},
  year={1993},
  publisher={Elsevier}
}

@article{shiferaw2019gaze,
  title={Gaze entropy measures reveal alcohol-induced visual scanning impairment during ascending and descending phases of intoxication},
  author={Shiferaw, Brook A and Downey, Luke A and Crewther, David},
  journal={Journal of Studies on Alcohol and Drugs},
  volume={80},
  number={2},
  pages={236--245},
  year={2019},
  publisher={Rutgers University}
}

@inproceedings{shic2008amorphous,
  title={The amorphous fixation measure revisited: With applications to autism},
  author={Shic, Frederick and Chawarska, Katarzyna and Scassellati, Brian},
  booktitle={Proceedings of the 30th Annual Meeting of the Cognitive Science Society},
  pages={2221--2226},
  year={2008},
  organization={Cognitive Science Society}
}

@article{sato2024gaze,
  title={Gaze transition entropy as a measure of attention allocation in a dynamic workspace involving automation},
  author={Sato, Toyomi and Cui, Renjie and others},
  journal={Scientific Reports},
  volume={14},
  number={1},
  pages={1--11},
  year={2024},
  publisher={Nature Publishing Group}
}

@article{russo1983strategies,
  title={Strategies for multiattribute binary choice},
  author={Russo, J Edward and Dosher, Barbara Anne},
  journal={Journal of Experimental Psychology: Learning, Memory, and Cognition},
  volume={9},
  number={4},
  pages={676},
  year={1983},
  publisher={American Psychological Association}
}

@article{goldberg1999computer,
  title={Computer interface evaluation using eye movements: methods and constructs},
  author={Goldberg, Joseph H and Kotval, Xerxes P},
  journal={International Journal of Industrial Ergonomics},
  volume={24},
  number={6},
  pages={631--645},
  year={1999},
  publisher={Elsevier}
}

@article{noton1971eye,
  title={Eye movements and visual perception},
  author={Noton, David and Stark, Lawrence},
  journal={Scientific American},
  volume={224},
  number={6},
  pages={34--43},
  year={1971},
  publisher={JSTOR}
}

@article{borys2017eye,
  title={Eye-tracking metrics in perception and visual attention research},
  author={Borys, Magdalena and Plechawska-W{\'o}jcik, Ma{\l}gorzata},
  journal={EJMT},
  volume={3},
  number={16},
  pages={11--23},
  year={2017}
}

@incollection{irwin2013fixation,
  title={Fixation location and fixation duration as indices of cognitive processing},
  author={Irwin, David E},
  booktitle={The interface of language, vision, and action},
  pages={105--133},
  year={2013},
  publisher={Psychology Press}
}

@article{canny1986computational,
  title = {A Computational Approach to Edge Detection},
  author = {Canny, John},
  journal = {IEEE Transactions on Pattern Analysis and Machine Intelligence},
  volume = {PAMI-8},
  number = {6},
  pages = {679--698},
  year = {1986},
  doi = {10.1109/TPAMI.1986.4767851}
}

@article{just1976eye,
  title   = {Eye fixations and cognitive processes},
  author  = {Just, Marcel Adam and Carpenter, Patricia A.},
  journal = {Cognitive Psychology},
  volume  = {8},
  number  = {4},
  pages   = {441--480},
  year    = {1976}
}

@article{mackworth1967gaze,
  title   = {The gaze selects informative details within pictures},
  author  = {Mackworth, Norman H. and Morandi, Anthony J.},
  journal = {Perception \& Psychophysics},
  volume  = {2},
  number  = {11},
  pages   = {547--552},
  year    = {1967}
}

@article{henderson1999high,
  title   = {High-level scene perception},
  author  = {Henderson, John M. and Hollingworth, Andrew},
  journal = {Annual Review of Psychology},
  volume  = {50},
  pages   = {243--271},
  year    = {1999}
}

@article{lan2026impact,
  title={The impact of task demand on subtitle reading in video games: an eye-tracking study},
  author={Lan, Haiting and Kruger, Jan-Louis and Richardson, Michael},
  journal={Applied Psycholinguistics},
  volume={47},
  pages={e10},
  year={2026},
  publisher={Cambridge University Press}
}

@inproceedings{blascheck2014state,
  title={State-of-the-art of visualization for eye tracking data.},
  author={Blascheck, Tanja and Kurzhals, Kuno and Raschke, Michael and Burch, Michael and Weiskopf, Daniel and Ertl, Thomas},
  booktitle={Eurovis (stars)},
  pages={29},
  year={2014}
}

\appendix
\section{AOI Template Examples}
\label{app:boundary}
\begin{figure}[t]
    \centering
    \includegraphics[width=0.7\linewidth]{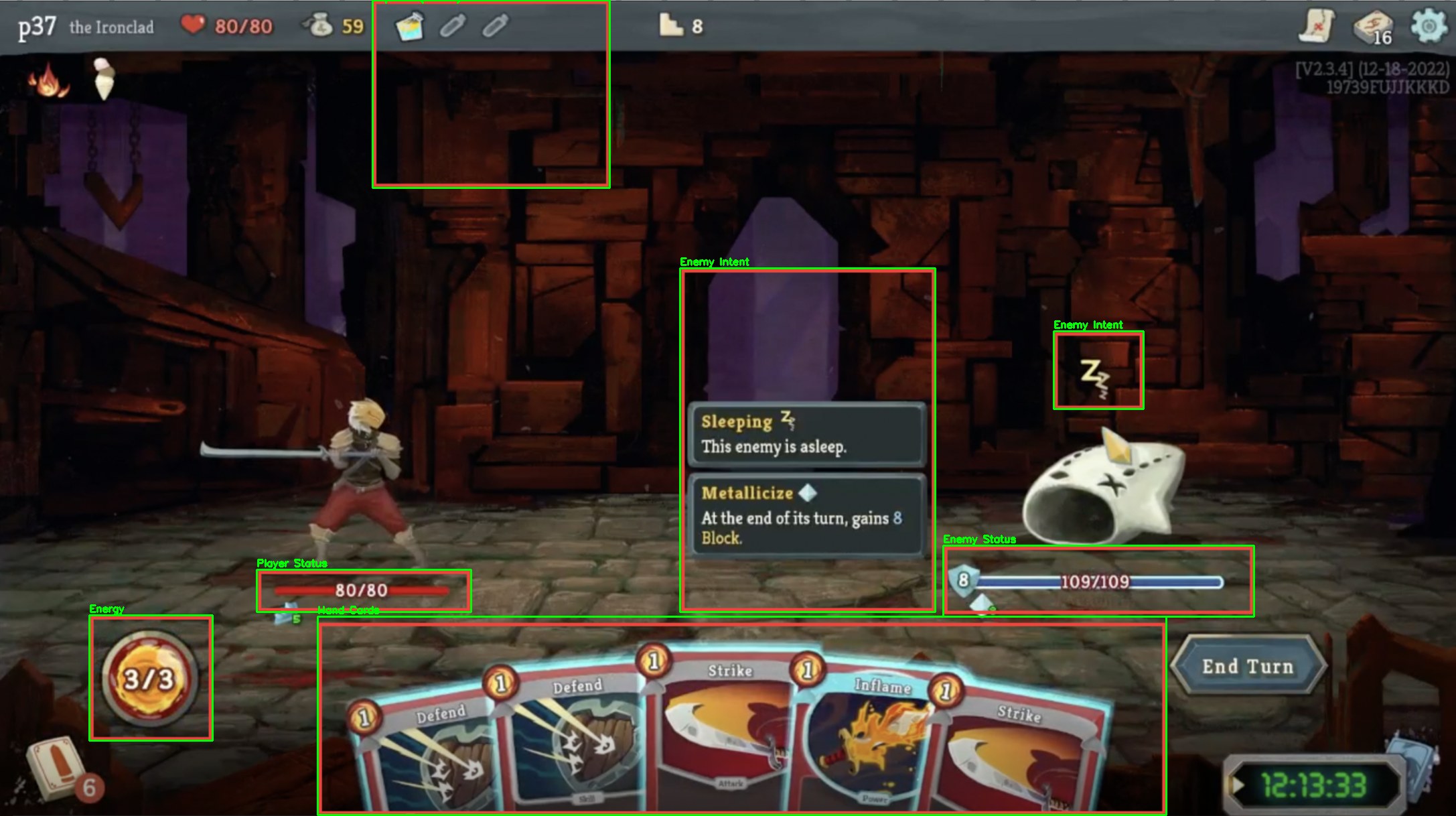}\\
    \vspace{0.5em}
    \includegraphics[width=0.7\linewidth]{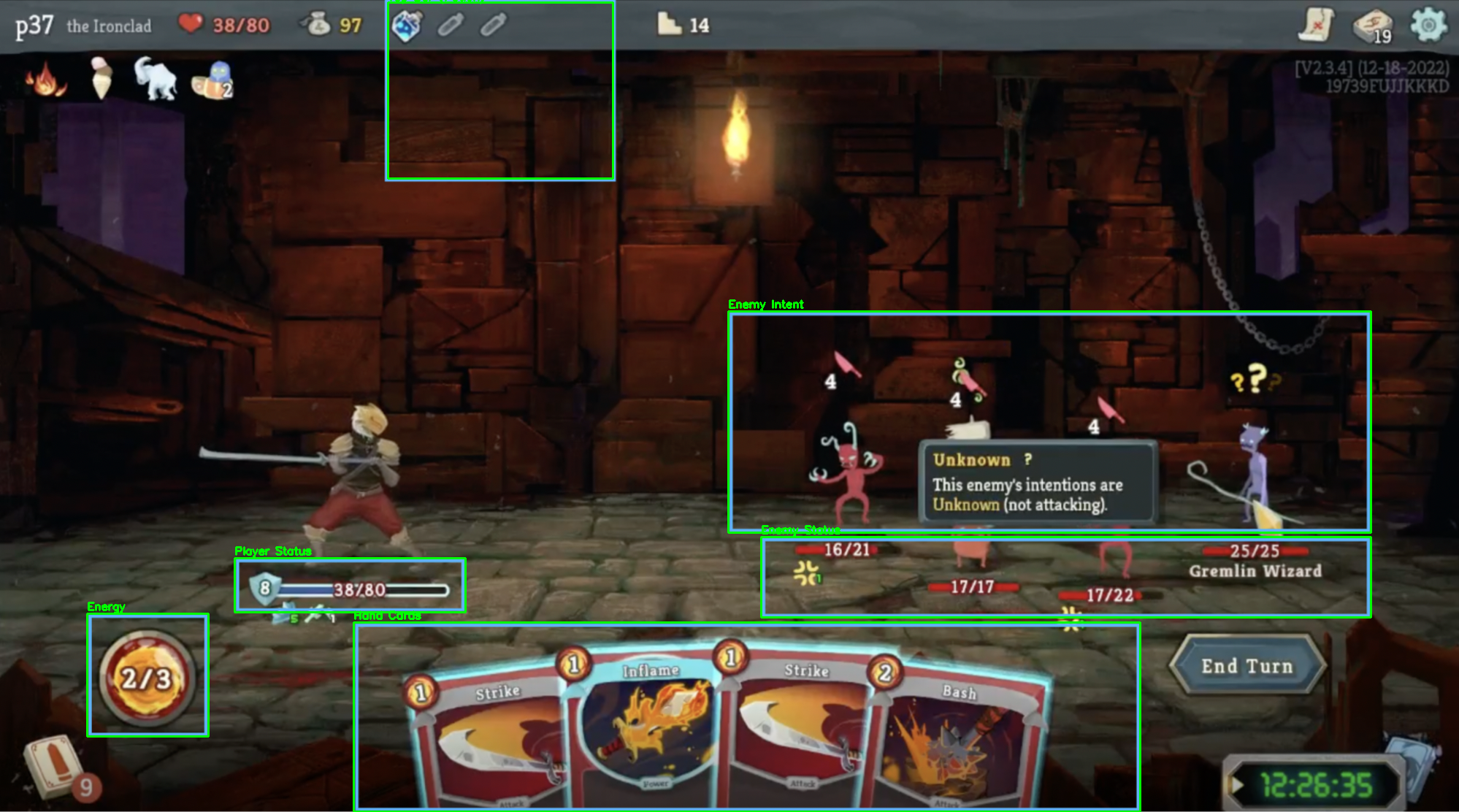}\\
    \vspace{0.5em}
    \includegraphics[width=0.7\linewidth]{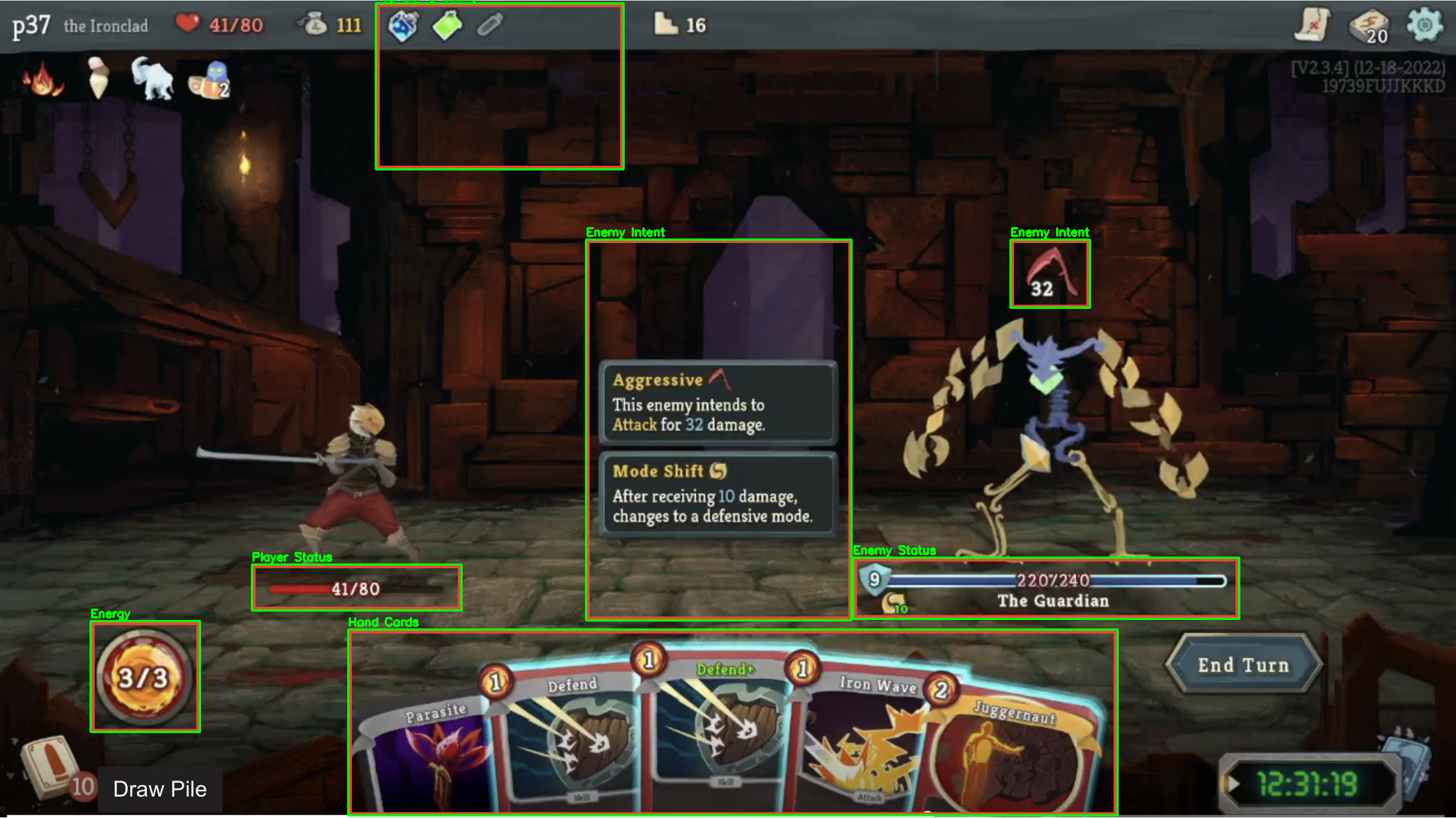}
    \caption{
    Example AOI templates used for gaze mapping across the three battle types.
    Top: single-enemy encounters.
    Middle: multiple-enemy encounters.
    Bottom: boss encounters.
    Coloured overlays indicate the six predefined AOIs (R1--R6): Enemy Intent, Hand Cards, Enemy Status, Player Status, Potion Panel, and Energy.
    }
    \label{fig:aoi_templates}
    \Description[]{}
\end{figure}

Figure~\ref{fig:aoi_templates} illustrates example AOI boundary templates used for gaze mapping across the three battle types.
AOIs were defined at the level of functional interface regions and localised using template-based layouts anchored to the game interface.
Separate templates were constructed for single-enemy, multiple-enemy, and boss encounters to account for structural differences in enemy-related interface elements.
These templates were reused across encounters of the same type to ensure consistent AOI definitions during gaze mapping.

\section{AOI Template Alignment Validation}
\label{app:aoi_validation}
Table~\ref{tab:aoi_validation} reports the per-AOI alignment results from the visual inspection of 156 frames sampled across battle types. Across battle types, within-template rates were 94.9\% for Boss combat, 96.4\% for Single-enemy combat, and 95.1\% for Multi-enemy combat. Misalignment was concentrated in the Enemy Intent AOI (R1), which exhibited the greatest positional variation across encounters due to differences in enemy number, positioning, and animation states. Non-diegetic UI elements with fixed on-screen positions (Player Status, Energy, Potion Panel, Hand Cards) showed consistent alignment across encounters. 

\begin{table}[h]
\centering
\caption{Per-AOI template alignment rates across 156 sampled frames.}
\label{tab:aoi_validation}
\begin{tabular}{lllc}
\toprule
AOI & Category & Within-template rate & Frames within / Total \\
\midrule
R4 Player Status & Player Information & 100\% & 156/156 \\
R6 Energy & Action Resources & 100\% & 156/156 \\
R5 Potion Panel & Auxiliary Resources & 99.4\% & 155/156 \\
R2 Hand Cards & Action Resources & 97.4\% & 152/156 \\
R3 Enemy Status & Enemy Information & 94.2\% & 147/156 \\
R1 Enemy Intent & Enemy Information & 82.1\% & 128/156 \\
\midrule
\textbf{Overall (all AOIs)} & & \textbf{95.4\%} & \textbf{893/936} \\
\bottomrule
\end{tabular}
\end{table}

\end{document}